\documentclass[aps,prb,reprint,superscriptaddress]{revtex4-2}

\usepackage{graphicx}
\usepackage{dcolumn}
\usepackage{bm}
\usepackage{amsmath}
\usepackage{amssymb}
\usepackage{mathtools}
\usepackage{lipsum}

\usepackage[utf8]{inputenc}
\usepackage{etoolbox}
\usepackage{soul}
\usepackage[colorlinks=true,citecolor=blue,urlcolor=blue]{hyperref}
\usepackage{tcolorbox}
\usepackage{siunitx}
\usepackage{braket}
\usepackage{todonotes}
\usepackage{array}
\usepackage{multirow}
\usepackage{float}
\usepackage{pifont}
\usepackage{lipsum}
\usepackage{mhchem}

\usepackage{newtxtext,newtxmath}
\DeclareMathAlphabet\mathbfcal{OMS}{cmsy}{b}{n}
\DeclareSymbolFontAlphabet{\mathcal}{symbols}
\DeclareMathAlphabet{\mathcal}{OMS}{cmsy}{m}{n}
\SetMathAlphabet{\mathcal}{bold}{OMS}{cmsy}{b}{n}

\newcommand{\iu}{\mathrm{i}}
\newcommand{\ec}{\mathrm{e}}

\newcommand{\XTM}{\vec{K}}
\newcommand{\XRM}{\vec{k}}
\newcommand{\XRMmagnitude}{k}
\newcommand{\BXTM}{\vec{Q}}
\newcommand{\BXRM}{\vec{q}}
\newcommand{\genSpinOne}{i}
\newcommand{\genSpinTwo}{j}
\newcommand{\excitonMass}{M_{\mathrm{X}}}
\renewcommand{\vec}[1]{\boldsymbol{#1}}
\allowdisplaybreaks

\definecolor{darkpastelgreen}{rgb}{0.01, 0.75, 0.24}

\newcommand{\Xmodeindex}{\lambda}
\newcommand{\XmodeindexNoSpin}{\tilde{\lambda}}
\newcommand{\BXmodeindex}{\eta}

\begin{document}

\title[]{Theory of exciton--exciton interactions with nontrivial electronic bands}
\author{L. Maisel Licerán}
\email{l.maiselliceran@uu.nl}
\affiliation{Institute for Theoretical Physics and Center for Extreme Matter and Emergent Phenomena, Utrecht University, Princetonplein 5, 3584 CC Utrecht, The Netherlands}
\author{H. T. C. Stoof}
\affiliation{Institute for Theoretical Physics and Center for Extreme Matter and Emergent Phenomena, Utrecht University, Princetonplein 5, 3584 CC Utrecht, The Netherlands}

\date{\today}

\begin{abstract}
    \noindent
	Excitons are bound electron--hole pairs in semiconductors whose interactions influence optical properties and determine their collective behavior and emergent correlated phases.
	Despite the theoretical and experimental significance of these phenomena, a microscopic theory of exciton--exciton interactions that takes into account the full composite nature of these quasiparticles in systems with an arbitrary single-particle Hamiltonian is lacking.
	Here, we present a general theory of exciton--exciton interactions that builds upon our previous publication [\href{https://doi.org/10.1103/hmkm-jxzl}{Physical Review B \textbf{113}, 155440 (2026)}], which applied only to conventional semiconductors with weak band hybridization.
	By contrast, the theory presented here is valid for virtually any semiconducting crystal and can be applied to the study of biexciton formation and correlated excitonic phases in systems where band topology, quantum geometry, and moiré effects play a significant role.
	In addition to the analytical theory, we also present results in simple flat-band models with a nontrivial quantum metric and show that the biexcitonic spectrum differs strongly from that in the case without quantum geometry.
	Our theory paves the way for accurate quantitative calculations of biexciton spectra and excitonic many-body phases in modern condensed-matter systems.
\end{abstract}

\maketitle

\section{Introduction}

Excitons are bound electron--hole pairs in semiconductors that were predicted almost a century ago by Frenkel \cite{frenkel1931transformation1,frenkel1931transformation2} and Wannier \cite{wannier1937structure}.
They play a crucial role in the optical properties of many systems and influence the behavior of optoelectronic devices such as solar cells \cite{zhang1998effects,yoo2007analysis,nozik2010semiconductor,dias2021excitonic,he2023exciton} and LEDs \cite{shuai1999singlet,ji2014efficient,hu2024exciton,baek2025exciton}.
In many systems of interest the excitons are of the Wannier type, meaning that their size is much larger than the underlying lattice spacing.
The properties of single Wannier excitons have been studied in detail and are well understood in a large variety of systems, out of which two-dimensional (2D) materials hold particular promise for future applications due to their unprecedented degree of control and tunability \cite{chernikov2014exciton,selig2016excitonic,wang2016radiative,zhang2017magnetic,xiao2017excitons,mueller2018exciton,rodin2020collective,lee2023recent,glazov2024excitons}.

Beyond the free exciton picture, the topic of exciton--exciton interactions is of great theoretical and experimental interest.
Not only do these influence the dynamics and photoluminescence spectrum of the system, but they also contribute to exciton--exciton annihilation and valley depolarization \cite{peyghambarian1984blue,kossacki2005exciton,stone2009exciton,sim2013exciton,mouri2014nonlinear,sun2014observation,kumar2014exciton,soavi2016exciton,dostal2018direct,mahmood2018observation,purz2021coherent,birkmeier2022probing,steinhoff2024exciton}.
Moreover, they dictate the properties of two-exciton bound states known as biexcitons.
Their formation and dynamics have been studied in different kinds of systems such as bulk semiconductors and epitaxial films \cite{mysyrowicz1968excitonic,grun1976polariton,schrey1978experimental,yamamoto2001biexciton,teke2004excitonic}, quantum wells \cite{miller1982biexcitons,yamada1994time,herz1997biexciton,ali2018evidence}, quantum dots and quantum wires \cite{banyai1987excitons,bacher1999biexciton,scheibner2007photoluminescence,vonk2021biexciton,sun2022biexciton,diroll2023long,huang2023nonlocal}, transition-metal dichalcogenides (TMDs) \cite{you2015observation,plechinger2015identification,pei2017excited,steinhoff2018biexciton,wang2019studying,conway2022direct,sharma2022engineering,jadczak2023biexciton}, excited semiconductor nanoplatelets \cite{kunneman2014nature,peng2021observation}, and perovskite nanocrystals \cite{thouin2018stable,cho2024size}.
Furthermore, excitons in modern condensed-matter systems such as TMDs can realize correlated phases of matter.
In this context, exciton condensation and superfluidity \cite{wang2019evidence,shi2022bilayer,cutshall2025imaging}, excitonic crystal ordering \cite{bai2023evidence}, excitonic \cite{ma2021strongly,chen2022excitonic,gu2022dipolar,zhang2022correlated} and Mott insulators \cite{xiong2023correlated,gao2024excitonic,lian2024valley}, exciton Wigner crystals \cite{qi2026exciton}, exciton density waves \cite{zeng2023exciton}, and ferroelectric ordering of excitons \cite{devenica2026collective} have all been reported in TMD systems.
In particular, the physics of flat-band moir\'e excitons in twisted TMD multilayers is dominated by their interactions, potentially allowing excitons to be used as a quantum simulator via the realization of a Bose--Hubbard model \cite{shimazaki2020strongly,kennes2021moire,gotting2022moire,park2023dipole}.
The properties of such correlated phases, as well as the phase diagrams of interacting excitons, significantly depend on the details of the underlying exciton--exciton interactions.

Despite its importance, the problem of excitonic interactions is complex due to the composite nature of these quasiparticles.
As each exciton is made up of an electron and a hole, the problem of two interacting excitons constitutes a complicated four-body problem.
Previous approaches either perform approximations that apply only to specific situations, such as a strong electron--hole mass imbalance \cite{Zimmermann_2007,Zimmermann_2008}, or do not fully take into account the different exchange processes that can take place between the two identical electrons and holes that make up the four-body system \cite{Hanamura_1974,Haug_1984,Inoue_1998,deLeon_2001,Okumura_2001,Rochat_2001,Thilagam_2001,Elistratov_2016,Berman_2017,Shahnazaryan_2017,Gribakin_2021,Katsch_2018}.
In Ref.\ \cite{noordman2025variational} we have developed a theory of exciton--exciton interactions that takes into account all possible scattering and exchange processes between the constituents, thereby amending previous theories that only consider these effects approximately.
However, the theory presented there is only valid in conventional semiconductors, where band hybridization is largely absent throughout the entire Brillouin zone.
It is thus necessary to generalize our theory to the case where the electronic degrees of freedom can depend on the momentum.
This is the case in a plethora of modern condensed-matter systems where topology, quantum geometry, and moiré effects play a role.
In this article we present such a generalized theory of exciton--exciton interactions which is applicable to virtually any crystalline semiconducting system.
We also perform numerical calculations of biexcitonic binding energies in some simple systems with quantum geometry and compare the results to the conventional case, indirectly showing the important effects of these properties on the exciton--exciton interactions.

The rest of this article is organized as follows.
In Sec.\ \ref{sec:fw} we present the underlying system Hamiltonian, review the eigenproblem for single excitons, and present a variational eigenproblem for biexcitons from where an effective exciton--exciton interaction matrix can be read off.
In Sec.\ \ref{sec:TBEq} we present numerical results on biexciton binding energies in a square-lattice tight-binding model, and compare the results with the continuum Heitler--London case discussed in Ref.\ \cite{noordman2025variational}.
In Sec.\ \ref{sec:fbbls} we compare the biexcitonic energy spectrum of the previous section with two different models with flat bands and nontrivial quantum geometry, showing the strong influence of these effects on the binding and stability of bound two-exciton states.
Finally, in Sec.\ \ref{sec:chXXint:summary} we summarize our work and give an outlook for future research.

\section{Theoretical framework}
\label{sec:fw}

In this section we outline the exciton eigenvalue problem and then derive an eigenvalue equation for the bound state of two excitons (i.e., a biexciton) from where an effective exciton--exciton potential can be read off.

\subsection{System Hamiltonian}
\label{sec:H}

We consider a $d$-dimensional crystalline system of interacting electrons described by the Hamiltonian
\begin{equation}
	\hat{H} = \hat{H}_{0} + \hat{V} .
\end{equation}
The free part reads
\begin{equation}
	\hat{H}_{0} = \sum_{\genSpinOne \genSpinTwo} \sum_{\vec{k}} \hat{\psi}^{\dagger}_{\genSpinOne \vec{k}} [H^{0}_{\genSpinOne \genSpinTwo}(\vec{k}) - \mu \delta_{\genSpinOne \genSpinTwo} ]\hat{\psi}^{\vphantom{\dagger}}_{\genSpinTwo \vec{k}} ,
\end{equation}
where $\hat{\psi}^{\dagger}$ and $\hat{\psi}$ are electronic creation and annihilation operators.
The indices $\genSpinOne$ and $\genSpinTwo$ label all combined single-particle degrees of freedom such as orbital, sublattice, layer, and spin.
The Hamiltonian $H_{0}(\vec{k})$ can be either a periodic Bloch Hamiltonian or a $\vec{k} \boldsymbol{\cdot} \vec{p}$ Hamiltonian, and describes the single-particle spectrum of the system under consideration.
We denote its eigenvalues and eigenstates by $\epsilon^{\alpha}_{\vec{k}}$ and $\vert u^{\alpha}_{\vec{k}} \rangle$, respectively, where $\alpha$ is a band index.
We will assume that the spectrum is fully gapped and that the chemical potential $\mu$ lies within the gap.

In these conditions, the system can host excitons formed by exciting an electron from the filled valence bands into the empty conduction bands and binding together the resulting electron--hole pair via the interaction $\hat{V}$.
This term describes the electrostatic electron--electron repulsion and is written as
\begin{equation}
\label{eq:electrostaticIntV}
	\hat{V} = \frac{1}{2\mathcal{V}} \sum_{\genSpinOne \genSpinTwo} \sum_{\vec{q}} V^{\genSpinOne \genSpinTwo}_{\vec{q}} \, \hat{n}_{\genSpinOne \vec{q}} \hat{n}_{\genSpinTwo, {-}\vec{q}} ,
\end{equation}
where normal ordering is implied.
Here, $\hat{n}_{\genSpinOne \vec{q}} = \sum_{\vec{k}} \hat{\psi}^{\dagger}_{\genSpinOne, \vec{k} + \vec{q}} \hat{\psi}^{\vphantom{\dagger}}_{\genSpinOne \vec{k}}$ is a momentum-space density operator, and $\mathcal{V}$ is the system volume in $d$ dimensions.
Furthermore, $V^{\genSpinOne \genSpinTwo}_{\vec{q}}$ is the (repulsive) electrostatic interaction potential in momentum space, whose precise form depends on the type of Hamiltonian $H_{0}(\vec{k})$:
\begin{itemize}
	\item For a $\vec{k} \boldsymbol{\cdot} \vec{p}$ Hamiltonian, $V^{\genSpinOne \genSpinTwo}_{\vec{q}} = \int \mathrm{d}^{d} r \, V_{\genSpinOne \genSpinTwo}(\vec{r}) \, \ec^{{-} \iu \vec{q} \boldsymbol{\cdot} \vec{r}}$, and $\vec{q}$ formally runs over the entire unbounded momentum space.
	This is simply the continuum Fourier transform of $V_{\genSpinOne \genSpinTwo}(\vec{r})$.
	In this case, umklapp processes with nonzero reciprocal lattice vector are neglected, which is justified for $|\vec{q}|$ small compared to the microscopic reciprocal lattice vectors.
	\item For a Bloch Hamiltonian in a localized basis, $V^{\genSpinOne \genSpinTwo}_{\vec{q}} = \Omega \sum_{\vec{R}} V_{\genSpinOne \genSpinTwo}(\vec{R}) \, \ec^{{-} \iu \vec{q} \boldsymbol{\cdot} \vec{R}}$, where $\Omega$ is the volume of the unit cell, the sum runs over all Bravais lattice vectors $\vec{R}$, and $\vec{q}$ belongs to the first Brillouin zone.
	This is a lattice Fourier transform, so $V^{\genSpinOne \genSpinTwo}_{\vec{q}} = V^{\genSpinOne \genSpinTwo}_{\vec{q} + \vec{G}}$ for any reciprocal lattice vector $\vec{G}$ \footnote{This periodicity holds in the periodic gauge, in which $\hat{\psi}_{\genSpinOne \vec{k}} = N^{-1/2} \sum_{\vec{R}} \ec^{{-} \iu \vec{k} \boldsymbol{\cdot} \vec{R}} \hat{\psi}_{\genSpinOne \vec{R}}$ involves only the Bravais lattice vector $\vec{R}$ and not the orbital position $\vec{R} + \boldsymbol{\tau}_{\genSpinOne}$.
	In a gauge where the phase is chosen as $\ec^{{-} \iu \vec{k} \boldsymbol{\cdot} (\vec{R} + \boldsymbol{\tau}_{\genSpinOne})}$, one finds $\hat{n}_{\genSpinOne, \vec{q} + \vec{G}} = \ec^{\iu \vec{G} \boldsymbol{\cdot} \boldsymbol{\tau}_{\genSpinOne}} \hat{n}_{\genSpinOne \vec{q}}$, and the compensating phases appear in $V^{\genSpinOne \genSpinTwo}_{\vec{q}}$.
	Of course, the operator $\hat{V}$ itself is the same in both gauges.}.
\end{itemize}

The terms with $\vec{R} = \vec{0}$ require care.
The self-interaction ($\genSpinOne = \genSpinTwo$ and $\vec{R} = \vec{0}$) is removed by the normal ordering implied in Eq.\ \eqref{eq:electrostaticIntV}.
For two distinct flavors on the same site (e.g., opposite spins), $V(\vec{0})$ may diverge, like in the case of the Coulomb potential.
For orbitals of finite extent it is a finite Hubbard-like repulsion $U$ set by the orbital size, and should be regularized accordingly or treated as an input parameter.

In writing Eq.\ \eqref{eq:electrostaticIntV} we have assumed that the ``flavors'' $\genSpinOne$ and $\genSpinTwo$ are conserved at the electrostatic interaction vertex.
This is exact for conserved labels such as spin, and often for layer degrees of freedom.
It also holds to very good approximation in a basis of well-localized Wannier orbitals, where it is the leading (density--density) contribution, while exchange, pair-hopping, and density-assisted hopping terms, which do not conserve the orbital labels, are neglected.
We will develop the theory under this assumption, but will see that the generalization to the case without vertex conservation, which is sometimes a convenient starting point (e.g., for moiré systems), is straightforward.
However, in that case $V$ is no longer periodic in $\vec{q}$ at fixed indices.

\subsection{Exciton eigenvalue problem}

We define an exciton creation operator as
\begin{equation}
\label{eq:XCreationOpGamma}
	\hat{X}^{\dagger}_{\Xmodeindex \vec{K}} = \frac{1}{\sqrt{\mathcal{V}}} \sum_{cv} \sum_{\XRM} \Phi^{cv}_{\Xmodeindex \vec{K}}(\vec{k}) \hat{\psi}^{\dagger}_{c, \vec{k} + \gamma \vec{K}} \hat{\psi}^{\vphantom{\dagger}}_{v, \vec{k} - \bar{\gamma} \vec{K}} ,
\end{equation}
where $\XTM$ is the total exciton momentum and $\Xmodeindex$ denotes a set of quantum numbers used to label the specific exciton state.
The sum runs over the relative momentum $\XRM$ and a set of conduction and valence bands $c$ and $v$ which are fully empty and filled, respectively.
The normalization of the exciton wave function is given by $\frac{1}{\mathcal{V}} \sum_{cv \vec{k}} |\Phi^{cv}_{\Xmodeindex \vec{K}}(\vec{k})|^{2} = 1$.

In writing down Eq.\ \eqref{eq:XCreationOpGamma} we have introduced a number $\gamma$ (and also $\bar{\gamma} \equiv 1 - \gamma$ for convenience) which can be freely chosen without modifying the total momentum $\vec{K}$ of the exciton.
In a $\vec{k} \boldsymbol{\cdot} \vec{p}$ model with parabolic bands, it is useful to choose $\gamma = m_{\mathrm{e}} / (m_{\mathrm{e}} + m_{\mathrm{h}})$, with $m_{\mathrm{e}}$ and $m_{\mathrm{h}}$ the masses of the electron and hole that make up the exciton, respectively.
This choice causes the kinetic portion of the exciton Hamiltonian to decouple into a relative and a center-of-mass part.
When an effective-mass description is not appropriate, it is customary to choose $\gamma = \bar{\gamma} = 1/2$, or $\gamma = 0$ or $1$.
The latter choice can be especially convenient in the case of numerical calculations in lattice models to avoid fractional prefactors leading to momenta lying oustide of the discretized grid.

Minimizing the expectation value of $\hat{H}$ in a one-exciton state with respect to $\Phi^{*}$ leads to the so-called Bethe--Salpeter equation (BSE) for the relative wave function,
\begin{equation}
\label{eq:excitonBSE}
	H^{\text{BS}}_{\XTM} \Phi^{\Xmodeindex}_{\XTM} = \varepsilon^{\Xmodeindex}_{\XTM} \Phi^{\Xmodeindex}_{\XTM} .
\end{equation}
The eigenvalues $\varepsilon^{\Xmodeindex}_{\XTM}$ of this equation are the exciton dispersions.
The components of the Bethe--Salpeter Hamiltonian in the band space read
\begin{equation}
	[H^{\text{BS}}]^{cvc'v'}_{\XTM \XRM \XRM'} = \delta_{\XRM \XRM'} \delta_{cc'} \delta_{vv'} \Delta^{cv}_{\XTM \XRM} - \mathcal{V}^{-1} [D - X]^{cvc'v'}_{\XTM \XRM \XRM'} ,
\end{equation}
where
\begin{equation}
	\Delta^{cv}_{\XTM \XRM} = \epsilon^{c}_{\vec{k} + \gamma \vec{K}} - \epsilon^{v}_{\vec{k} - \bar{\gamma} \vec{K}} .
\end{equation}
The so-called direct and exchange interactions read
\begin{subequations}
\label{eq:DXboth}
	\begin{align}
		&D^{cvc'v'}_{\vec{K} \vec{k} \vec{k}'} = \sum_{\genSpinOne \genSpinTwo} \big(u^{c}_{\vec{k} + \gamma \vec{K}} \big)^{*}_{\genSpinOne} \big(u^{c'}_{\vec{k}' + \gamma \vec{K}}\big)^{\vphantom{*}}_{\genSpinOne} V^{\genSpinOne \genSpinTwo}_{\vec{k} - \vec{k}'} \big(u^{v'}_{\vec{k}' - \bar{\gamma} \vec{K}} \big)^{*}_{\genSpinTwo} \big(u^{v}_{\vec{k} - \bar{\gamma} \vec{K}} \big)^{\vphantom{*}}_{\genSpinTwo} , \\
		&X^{cvc'v'}_{\vec{K} \vec{k} \vec{k}'} = \sum_{\genSpinOne \genSpinTwo} \big( u^{c}_{\vec{k} + \gamma \vec{K}} \big)^{*}_{\genSpinOne} \big( u^{v}_{\vec{k} - \bar{\gamma} \vec{K}} \big)^{\vphantom{*}}_{\genSpinOne} V^{\genSpinOne \genSpinTwo}_{\vec{K}} \big( u^{v'}_{\vec{k}' - \bar{\gamma} \vec{K}} \big)^{*}_{\genSpinTwo} \big( u^{c'}_{\vec{k}' + \gamma \vec{K}} \big)^{\vphantom{*}}_{\genSpinTwo} ,
	\end{align}
\end{subequations}
where $(u^{\alpha}_{\vec{k}})^{\vphantom{*}}_{\genSpinOne} \equiv \langle \genSpinOne \vert u^{\alpha}_{\vec{k}} \rangle$ is the $\genSpinOne$ component of the eigenstate $\vert u^{\alpha}_{\vec{k}} \rangle$.
The form factors accompanying the electrostatic interaction $V^{\genSpinOne \genSpinTwo}_{\vec{q}}$ arise because the electronic operators in Eq.\ \eqref{eq:electrostaticIntV} must be transformed to the band space in which the exciton creation operator of Eq.\ \eqref{eq:XCreationOpGamma} is expressed.
In systems where the single-particle model features nontrivial band geometry or topology, the dressing of the electrostatic interaction via these form factors is crucial to obtain the correct exciton spectrum.

We note that the solutions to Eq.\ \eqref{eq:excitonBSE} in general contain both bound states and unbound or scattering states, but we refer as ``excitons'' to the bound states only.
The relative wave functions satisfy the completeness relations
\begin{subequations}
	\begin{align}
		\frac{1}{\mathcal{V}} \sum_{cv\XRM} [\Phi^{cv}_{\Xmodeindex \XTM}(\XRM)]^{*} \Phi^{cv}_{\Xmodeindex' \XTM}(\XRM) &= \delta_{\Xmodeindex \Xmodeindex'} , \\
		\sum_{\Xmodeindex} [\Phi^{cv}_{\Xmodeindex \XTM}(\XRM)]^{*} \Phi^{c'v'}_{\Xmodeindex \XTM}(\XRM') &= \mathcal{V} \delta_{\XRM \XRM'} \delta_{cc'} \delta_{vv'} ,
	\end{align}
\end{subequations}
where the sum over $\Xmodeindex$ in the second line must run over \emph{all} electron--hole states (i.e., bound \emph{and} unbound) for the equality to hold.

\subsection{General biexciton eigenvalue problem}
\label{sec:genBXProblem}

Now we consider the biexciton problem, i.e., the problem of a bound state of two excitons.
To begin, we define the biexciton creation operator for a biexciton state $\BXmodeindex$ with total momentum $\vec{Q}$ as
\begin{equation}
\label{eq:biexcitonCreationOperatorGamma}
	\hat{B}^{\dagger}_{\BXmodeindex \vec{Q}} = \frac{1}{2 \sqrt{\mathcal{V}}} \sum_{\Xmodeindex_{1} \Xmodeindex_{2} \vec{q}} \Psi^{\Xmodeindex_{1} \Xmodeindex_{2}}_{\BXmodeindex \vec{Q}}(\vec{q}) \hat{X}^{\dagger}_{\Xmodeindex_{1}, \Gamma \vec{Q} + \vec{q}} \hat{X}^{\dagger}_{\Xmodeindex_{2}, \bar{\Gamma} \vec{Q} - \vec{q}} ,
\end{equation}
where $\Psi$ is the relative biexciton wave function and the factor of 2 in the normalization is merely for convenience.
Again, $\bar{\Gamma} \equiv 1 - \Gamma$ and $\Gamma$ can be freely chosen without changing the total momentum of the biexciton.
For further convenience we also define
\begin{subequations}
	\begin{align}
		g &= 2 \gamma - 1 , \\
		G &= 2 \Gamma - 1 .
	\end{align}
\end{subequations}
At this stage, the sum over exciton states $\Xmodeindex_{1}$ and $\Xmodeindex_{2}$ in Eq.\ \eqref{eq:biexcitonCreationOperatorGamma} can run over the entirety of the electron--hole space or over a restricted set of states only.
Because of the commuting nature of the exciton creation operators, the coefficients can be chosen to satisfy
\begin{equation}
\label{eq:BXWFExchangeSymmGamma}
	\Psi^{\Xmodeindex_{1} \Xmodeindex_{2}}_{\BXmodeindex \vec{Q}}(\vec{q}) = \Psi^{\Xmodeindex_{2} \Xmodeindex_{1}}_{\BXmodeindex \vec{Q}} ({-}\vec{q} - G \vec{Q}) .
\end{equation}
This simply expresses the symmetry of the biexciton wave function under exciton exchange.
Note that the case $\Gamma = 1/2$, which is especially appropriate when both excitons have the same mass, results in $G = 0$ and thus $\Psi^{\Xmodeindex_{1} \Xmodeindex_{2}}_{\BXmodeindex \vec{Q}}(\vec{q}) = \Psi^{\Xmodeindex_{2} \Xmodeindex_{1}}_{\BXmodeindex \vec{Q}} ({-}\vec{q})$ as in our Ref.\ \cite{noordman2025variational}.

Now we minimize the expectation of $\hat{H}$ in a one-biexciton state with respect to $\Psi^{*}$ and obtain the \emph{generalized} eigenvalue problem
\begin{equation}
\label{eq:BXGenEVProb}
	\mathcal{H}^{\vphantom{\BXmodeindex}}_{\BXTM} \Psi^{\BXmodeindex}_{\BXTM} = \mathcal{E}_{\BXTM}^{\BXmodeindex} \mathcal{A}^{\vphantom{\BXmodeindex}}_{\BXTM} \Psi^{\BXmodeindex}_{\BXTM} .
\end{equation}
Here, $\mathcal{E}^{\BXmodeindex}_{\BXTM}$ is the biexciton dispersion, and the biexciton ``Hamiltonian'' has components \footnote{Note that $\mathcal{H}_{\BXTM}$ as defined here has dimensions of energy times volume, but all factors of $\mathcal{V}$ in the generalized eigenproblem are correctly compensated at both sides of Eq.\ \eqref{eq:BXGenEVProb}.}
\begin{equation}
\label{eq:BXHamiltonian}
	\begin{split}
		& \mathcal{H}^{\Xmodeindex_{1} \Xmodeindex_{2} \Xmodeindex'_{1} \Xmodeindex'_{2}}_{\BXTM \BXRM \BXRM'} = \frac{1}{2} E^{\Xmodeindex_{1} \Xmodeindex_{2} \Xmodeindex'_{1} \Xmodeindex'_{2}}_{\BXTM \BXRM \BXRM'} \mathcal{A}^{\Xmodeindex_{1} \Xmodeindex_{2} \Xmodeindex'_{1} \Xmodeindex'_{2}}_{\BXTM \BXRM \BXRM'} + \frac{1}{2} \mathcal{U}^{\Xmodeindex_{1} \Xmodeindex_{2} \Xmodeindex'_{1} \Xmodeindex'_{2}}_{\BXTM \BXRM \BXRM'} ,
	\end{split}
\end{equation}
where
\begin{equation}
	E^{\Xmodeindex_{1} \Xmodeindex_{2} \Xmodeindex'_{1} \Xmodeindex'_{2}}_{\BXTM \BXRM \BXRM'} = \varepsilon^{\Xmodeindex_{1}}_{\Gamma \BXTM + \BXRM} + \varepsilon^{\Xmodeindex_{2}}_{\bar{\Gamma} \BXTM - \BXRM} + \varepsilon^{\Xmodeindex_{1}'}_{\Gamma \BXTM + \BXRM'} + \varepsilon^{\Xmodeindex_{2}'}_{\bar{\Gamma} \BXTM - \BXRM'}
\end{equation}
and the components of $\mathcal{U}$ are detailed further below.

The biexciton eigenproblem depends on an operator $\mathcal{A}$ whose matrix elements are given by
\begin{widetext}
	\begin{equation}
		\mathcal{A}^{\Xmodeindex_{1} \Xmodeindex_{2} \Xmodeindex'_{1} \Xmodeindex'_{2}}_{\BXTM \BXRM \BXRM'}
		= \frac{1}{4} \Big\{
		\mathcal{V} \delta_{\BXRM \BXRM'} \delta_{\Xmodeindex_{1} \Xmodeindex'_{1}} \delta_{\Xmodeindex_{2} \Xmodeindex'_{2}} 
		- [\mathcal{K}^{\mathrm{e}}]^{\Xmodeindex_{1} \Xmodeindex_{2} \Xmodeindex'_{1} \Xmodeindex'_{2}}_{\BXTM \BXRM \BXRM'}
		- [\mathcal{K}^{\mathrm{h}}]^{\Xmodeindex_{1} \Xmodeindex_{2} \Xmodeindex'_{1} \Xmodeindex'_{2}}_{\BXTM \BXRM \BXRM'}
		+ \mathcal{V} \delta_{\BXRM, {-} \BXRM' - G \BXTM} \delta_{\Xmodeindex_{1} \Xmodeindex'_{2}} \delta_{\Xmodeindex_{2} \Xmodeindex'_{1}} \Big\} .
	\end{equation}
	As explained in Ref.\ \cite{noordman2025variational}, this operator implements the antisymmetry under exchange of each pair of identical particles within the biexciton, and therefore we call it an antisymmetrizer.
	It contains electron- and hole-exchange overlap integrals which read
	\begin{subequations}
		\begin{align}
			\label{eq:Kc}
			\begin{split}
				&[\mathcal{K}^{\mathrm{e}}]^{\Xmodeindex_{1} \Xmodeindex_{2} \Xmodeindex'_{1} \Xmodeindex'_{2}}_{\BXTM \BXRM \BXRM'} \\
				&= \frac{1}{\mathcal{V}} \sum_{\vec{k}} \sum_{\text{bands}}
				[\Phi^{cv}_{\Xmodeindex_{1}, \Gamma \BXTM + \BXRM}(\vec{k})]^{*}
				[\Phi^{c'v'}_{\Xmodeindex_{2}, \bar{\Gamma} \BXTM - \BXRM}(\XRM + g \BXRM + \BXRM' + \gamma G \BXTM)]^{*}
				\Phi^{c'v}_{\Xmodeindex_{1}', \Gamma \BXTM + \BXRM'}(\XRM - \bar{\gamma}(\BXRM - \BXRM'))
				\Phi^{cv'}_{\Xmodeindex_{2}', \bar{\Gamma} \BXTM - \BXRM'}(\XRM + \gamma (\BXRM + \BXRM' + G \BXTM)) ,
			\end{split}
			\\
			\label{eq:Kv}
			\begin{split}
				&[\mathcal{K}^{\mathrm{h}}]^{\Xmodeindex_{1} \Xmodeindex_{2} \Xmodeindex'_{1} \Xmodeindex'_{2}}_{\BXTM \BXRM \BXRM'} = \\
				&= \frac{1}{\mathcal{V}} \sum_{\XRM} \sum_{\text{bands}}
				[\Phi^{cv}_{\Xmodeindex_{1}, \Gamma \BXTM + \BXRM}(\XRM)]^{*}
				[\Phi^{c'v'}_{\Xmodeindex_{2}, \bar{\Gamma} \BXTM - \BXRM}(\XRM + g \BXRM - \BXRM' - \bar{\gamma} G \BXTM)]^{*}
				\Phi^{cv'}_{\smash[b]{\Xmodeindex'_{\smash[t]{1}}}, \Gamma \BXTM + \BXRM'}(\XRM + \gamma (\BXRM - \BXRM'))
				\Phi^{c'v}_{\smash[b]{\Xmodeindex'_{\smash[t]{2}}}, \bar{\Gamma} \BXTM - \BXRM'}(\XRM - \bar{\gamma} (\BXRM + \BXRM' + G \BXTM)) .
			\end{split}
		\end{align}
	\end{subequations}
\end{widetext}
Here and below, the sum over bands runs over all conduction bands ($c, c', \dots$) and valence bands ($v, v', \dots$) appearing in the expression.
The exchange overlap integrals satisfy the relations
\begin{subequations}
	\begin{align}
	\label{eq:KetoKhfirst}
		[\mathcal{K}^{\mathrm{e}}]^{\Xmodeindex_{1} \Xmodeindex_{2} \Xmodeindex'_{1} \Xmodeindex'_{2}}_{\BXTM \BXRM \BXRM'}
		&= [\mathcal{K}^{\mathrm{h}}]^{\Xmodeindex_{1} \Xmodeindex_{2} \Xmodeindex'_{2} \Xmodeindex'_{1}}_{\BXTM \BXRM , {-}\BXRM' - G \BXTM} \\
		&= [\mathcal{K}^{\mathrm{h}}]^{\Xmodeindex_{2} \Xmodeindex_{1} \Xmodeindex'_{1} \Xmodeindex'_{2}}_{\BXTM, {-}\BXRM - G \BXTM, \BXRM'} \\
		&= [\mathcal{K}^{\mathrm{e}}]^{\Xmodeindex_{2} \Xmodeindex_{1} \Xmodeindex'_{2} \Xmodeindex'_{1}}_{\BXTM, {-}\BXRM - G \BXTM, {-}\BXRM' - G \BXTM} .
	\end{align}
\end{subequations}

Finally, the term $\mathcal{U}$ in Eq.\ \eqref{eq:BXHamiltonian} is an exciton--exciton interaction that has the structure
\begin{equation}
\label{eq:XXintUAllTerms}
	\mathcal{U}^{\Xmodeindex_{1} \Xmodeindex_{2} \Xmodeindex'_{1} \Xmodeindex'_{2}}_{\BXTM \BXRM \BXRM'} = \frac{1}{2} [\mathcal{U}^{0} + \mathcal{U}^{\mathrm{e}} + \mathcal{U}^{\mathrm{h}} + \mathcal{U}^{\mathrm{X}}]^{\Xmodeindex_{1} \Xmodeindex_{2} \Xmodeindex'_{1} \Xmodeindex'_{2}}_{\BXTM \BXRM \BXRM'} .
\end{equation}
Each of these terms contains electron--electron, hole--hole, and electron--hole scattering diagrams.
They describe scattering process with no particle exchange ($\mathcal{U}^{0}$), electron exchange ($\mathcal{U}^{\mathrm{e}}$), hole exchange ($\mathcal{U}^{\mathrm{h}}$), and combined electron--hole or exciton exchange ($\mathcal{U}^{\mathrm{X}}$).
Their explicit expressions are given in Appendix \ref{app:InteractionTermsExpressions} and involve the underlying excitonic wave functions, as well as the details of the single--particle physics owing to the presence of form factors as in the direct and exchange diagrams of the single-exciton problem.
The expressions presented in this Appendix assume that the electronic flavor is conserved at the electrostatic vertex, as explained at the end of Sec.\ \ref{sec:H}.
However, they can be easily modified to suit the general case in which the $\genSpinOne, \genSpinTwo, \dots$ labels can be different before and after the scattering, thus making the theory completely general.
This is clarified in Appendix\ \ref{app:generalVertex}.

The results reported here constitute a generalization to our previous theory in Ref.\ \cite{noordman2025variational}.
The crucial difference lies in the presence of single-particle form factors in the matrix elements of the exciton--exciton interaction $\mathcal{U}$.
As a result of these, the interaction $\mathcal{U}$ directly encodes effects owing to potential nontrivial topology and quantum geometry of the underlying free electrons.
Moreover, the matrix elements depend also on the relative exciton wave functions $\Phi$, whose zeros as a function of the total momentum at some fixed but arbitrary relative momentum can encode additional topological properties arising from the interactions \cite{paiva2024shift}.
Hence, band-geometrical and band-topological effects of both free particles and bound electron--hole pairs directly influence the biexciton spectrum.
We also note that the direct electron--hole scattering diagrams ($\mathcal{U}^{0}_{\mathrm{eh}}$ and $\mathcal{U}^{0}_{\mathrm{he}}$ in Appendix\ \ref{app:InteractionTermsExpressions}) now contain a repulsive ``exchange'' part arising from the nonzero overlap between conduction and valence Bloch states at different momenta, similar to what happens in the single-exciton problem.
All in all, while the results of Ref.\ \cite{noordman2025variational} are applicable to systems where band-geometrical, band-topological, and moir\'e effects are not present or negligible, the theory in this article is applicable to crystalline systems with arbitrary underlying single-particle Hamiltonians.

The generalized eigenvalue problem as written in Eq.\ \eqref{eq:BXGenEVProb} is not practical as it stands.
To see this, let $P_{\BXTM}$ be the exchange operator, i.e., $[P_{\BXTM} \Psi_{\BXTM}]^{\Xmodeindex_{1} \Xmodeindex_{2}}(\BXRM) = \Psi_{\BXTM}^{\Xmodeindex_{2} \Xmodeindex_{1}}({-}\BXRM - G \BXTM)$, and $S_{\BXTM} = (1 + P_{\BXTM})/2$.
Furthermore, let us define
\begin{equation}
\label{eq:BXHamiltonianReduced}
	\begin{split}
		h^{\Xmodeindex_{1} \Xmodeindex_{2} \Xmodeindex'_{1} \Xmodeindex'_{2}}_{\BXTM \BXRM \BXRM'} &= \frac{1}{2} E^{\Xmodeindex_{1} \Xmodeindex_{2} \Xmodeindex'_{1} \Xmodeindex'_{2}}_{\BXTM \BXRM \BXRM'} R^{\Xmodeindex_{1} \Xmodeindex_{2} \Xmodeindex'_{1} \Xmodeindex'_{2}}_{\BXTM \BXRM \BXRM'} + [\mathcal{U}^{0} + \mathcal{U}^{\mathrm{e}}]^{\Xmodeindex_{1} \Xmodeindex_{2} \Xmodeindex'_{1} \Xmodeindex'_{2}}_{\BXTM \BXRM \BXRM'}
	\end{split}
\end{equation}
and
\begin{equation}
	R^{\Xmodeindex_{1} \Xmodeindex_{2} \Xmodeindex'_{1} \Xmodeindex'_{2}}_{\BXTM \BXRM \BXRM'} = \mathcal{V} \delta_{\BXRM \BXRM'} \delta_{\Xmodeindex_{1} \Xmodeindex'_{1}} \delta_{\Xmodeindex_{2} \Xmodeindex'_{2}} - [\mathcal{K}^{\mathrm{e}}]^{\Xmodeindex_{1} \Xmodeindex_{2} \Xmodeindex'_{1} \Xmodeindex'_{2}}_{\BXTM \BXRM \BXRM'} ,
\end{equation}
which are the parts of $\mathcal{H}$ and $\mathcal{A}$ that only incorporate the direct and electron-exchange processes.
Using Eq.\ \eqref{eq:KetoKhfirst} and the properties of Eq.\ \eqref{eq:Urelations} satisfied by the different terms within $\mathcal{U}$, together with the invariance of $E$ under the exchange of the primed labels, one finds $\mathcal{H} = \frac{1}{2} h S$ and $\mathcal{A} = \frac{1}{2} R S$.
Hence, both operators annihilate all antisymmetric functions, and thus the eigenvalue problem of Eq.\ \eqref{eq:BXGenEVProb} is singular.
The physical solutions are those obeying Eq.\ \eqref{eq:BXWFExchangeSymmGamma}, i.e., $S \Psi = \Psi$, for which Eq.\ \eqref{eq:BXGenEVProb} reduces to
\begin{equation}
\label{eq:BXGenEVProbReduced}
	h^{\vphantom{\BXmodeindex}}_{\BXTM} \Psi^{\BXmodeindex}_{\BXTM} = \mathcal{E}^{\BXmodeindex}_{\BXTM} R^{\vphantom{\BXmodeindex}}_{\BXTM} \Psi^{\BXmodeindex}_{\BXTM} .
\end{equation}

In this way, the generalized eigenproblem of Eq.\ \eqref{eq:BXGenEVProb} has been reduced to a significantly simpler form.
The eigenvalue problem as written in Eq.\ \eqref{eq:BXGenEVProbReduced} does not involve $\mathcal{U}^{\mathrm{h}}$, $\mathcal{U}^{\mathrm{X}}$, or $\mathcal{K}^{\mathrm{h}}$, although all these exchange processes are still implicitly taken into account.
More importantly, the reduced eigenvalue problem is not singular, and the physical solutions can be straightforwardly recovered from its spectrum.
Since $\mathcal{H}$, $\mathcal{A}$, $h$, and $R$ are Hermitian, $h$ and $R$ commute with $P$, and thus Eq.\ \eqref{eq:BXGenEVProbReduced} is closed on the symmetric subspace.
Its symmetric solutions are precisely the physical solutions of the original eigenvalue problem of Eq.\ \eqref{eq:BXGenEVProb}, that is, those satisfying the symmetry requirement of Eq.\ \eqref{eq:BXWFExchangeSymmGamma}.
Meanwhile, the antisymmetric solutions of Eq.\ \eqref{eq:BXGenEVProbReduced} are unphysical and and must be discarded.

We also draw attention to the normalization of the biexciton wave function $\Psi$.
Due to the fact that the biexciton problem is a generalized eigenproblem, the wave functions are $\mathcal{A}$-orthonormal, and we can choose $[\Psi^{\BXmodeindex}_{\BXTM}]^{\dagger} \mathcal{A}^{\vphantom{\BXmodeindex}}_{\BXTM} \Psi^{\BXmodeindex'}_{\BXTM} = \delta_{\BXmodeindex \BXmodeindex'}$.
This is the correct convention when the factor of $2$ is included in the biexciton creation operator of Eq.\ \eqref{eq:biexcitonCreationOperatorGamma} and ensures that the biexciton states $\vert \BXmodeindex \BXTM \rangle \equiv B^{\dagger}_{\BXmodeindex \BXTM} \vert 0 \rangle$ are orthonormal for each $\BXTM$, that is, $\langle \BXmodeindex \BXTM \vert \BXmodeindex' \BXTM \rangle = \delta_{\BXmodeindex \BXmodeindex'}$.
In terms of $R$, the correct normalization of the biexciton wave function reads $\textstyle{\frac{1}{2}} [\Psi^{\BXmodeindex}_{\BXTM}]^{\dagger} R^{\vphantom{\BXmodeindex}}_{\BXTM} \Psi^{\BXmodeindex'}_{\BXTM} = \delta_{\BXmodeindex \BXmodeindex'}$, where the origin of the factor of $\textstyle{\frac{1}{2}}$ is easy to see by recalling that $S \Psi = \Psi$ for the physical states and that $\mathcal{A} = \textstyle{\frac{1}{2}} R S$.
Correctly normalizing the biexciton wave function is essential if one wishes to calculate properties such as the biexciton radius or oscillator strength.

All manipulations up to this point are possible no matter whether the exciton labels in Eq.\ \eqref{eq:biexcitonCreationOperatorGamma} run over all electron--hole states or only over a restricted subspace.
However, the next step is only possible in the latter case.
When the summation is restricted, the operator $R$ has an inverse, which can be written as the Dyson series
\begin{equation}
	\begin{split}
		& [R^{-1}]^{\Xmodeindex_{1} \Xmodeindex_{2} \Xmodeindex'_{1} \Xmodeindex'_{2}}_{\BXTM \BXRM \BXRM'} \\
		&= \mathcal{V} \delta_{\vec{q} \vec{q}'} \delta_{\Xmodeindex_{1} \Xmodeindex_{1}'} \delta_{\Xmodeindex_{2} \Xmodeindex'_{2}} + \frac{1}{\mathcal{V}} \sum_{\bar{\Xmodeindex}_{1} \bar{\Xmodeindex}_{2} \bar{\BXRM}} [R^{-1}]^{\Xmodeindex_{1} \Xmodeindex_{2} \bar{\Xmodeindex}_{1} \bar{\Xmodeindex}_{2}}_{\BXTM \BXRM \bar{\BXRM}} [\mathcal{K}^{\mathrm{e}}]^{\bar{\Xmodeindex}_{1} \bar{\Xmodeindex}_{2} \Xmodeindex'_{1} \Xmodeindex'_{2}}_{\BXTM \bar{\BXRM} \BXRM'} .
	\end{split}
\end{equation}
As shown in our Ref.\ \cite{noordman2025variational}, the reason the summation needs to be restricted to a subset of exciton states is that otherwise $R$ is proportional to a projector, which is not invertible.
By contrast, when the sum is restricted we can multiply Eq.\ \eqref{eq:BXGenEVProbReduced} from the left by $R^{-1}$ and rewrite the reduced biexciton eigenvalue problem as
\begin{widetext}
	\begin{equation}
	\label{eq:2XSchrodingerEquation}
		\big(\varepsilon^{\Xmodeindex_{1}}_{\Gamma \BXTM + \BXRM} + \varepsilon^{\Xmodeindex_{2}}_{\bar{\Gamma} \BXTM - \BXRM}\big) \Psi^{\Xmodeindex_{1} \Xmodeindex_{2}}_{\BXmodeindex \BXTM}(\BXRM) + \frac{1}{\mathcal{V}} \sum_{\Xmodeindex'_{1} \Xmodeindex'_{2} \BXRM'} [V^{\text{eff}}]^{\Xmodeindex_{1} \Xmodeindex_{2} \Xmodeindex'_{1} \Xmodeindex'_{2}}_{\BXTM \BXRM \BXRM'} \Psi^{\Xmodeindex'_{1} \Xmodeindex'_{2}}_{\BXmodeindex \BXTM}(\BXRM') = \mathcal{E}^{\BXmodeindex}_{\BXTM} \Psi^{\Xmodeindex_{1} \Xmodeindex_{2}}_{\BXmodeindex \BXTM}(\BXRM) .
	\end{equation}
	This has the form of a Schr\"odinger equation with a kinetic and potential term, with the matrix elements of the resulting effective exciton--exciton potential given by
	\begin{equation}
	\label{eq:XXVeff}
		\begin{split}
			[V^{\text{eff}}]^{\Xmodeindex_{1} \Xmodeindex_{2} \Xmodeindex'_{1} \Xmodeindex'_{2}}_{\BXTM \BXRM \BXRM'} = {-}\frac{1}{2 \mathcal{V}} &\sum_{\bar{\Xmodeindex}_{1} \bar{\Xmodeindex}_{2} \bar{\BXRM}} [R^{-1}]^{\Xmodeindex_{1} \Xmodeindex_{2} \bar{\Xmodeindex}_{1} \bar{\Xmodeindex}_{2}}_{\BXTM \BXRM \bar{\BXRM}} \big(\varepsilon^{\bar{\Xmodeindex}_{1}}_{\Gamma\BXTM + \bar{\BXRM}} + \varepsilon^{\bar{\Xmodeindex}_{2}}_{\bar{\Gamma}\BXTM - \bar{\BXRM}} - \varepsilon^{\Xmodeindex'_{1}}_{\Gamma\BXTM + \BXRM'} - \varepsilon^{\Xmodeindex'_{2}}_{\bar{\Gamma}\BXTM - \BXRM'}\big) [\mathcal{K}^{\mathrm{e}}]^{\bar{\Xmodeindex}_{1} \bar{\Xmodeindex}_{2} \Xmodeindex'_{1} \Xmodeindex'_{2}}_{\BXTM \bar{\BXRM} \BXRM'} \\
			+ \, \frac{1}{\mathcal{V}} &\sum_{\bar{\Xmodeindex}_{1} \bar{\Xmodeindex}_{2} \bar{\vec{q}}} [R^{-1}]^{\Xmodeindex_{1} \Xmodeindex_{2} \bar{\Xmodeindex}_{1} \bar{\Xmodeindex}_{2}}_{\BXTM \BXRM \bar{\BXRM}} [\mathcal{U}^{0} + \mathcal{U}^{\mathrm{e}}]^{\bar{\Xmodeindex}_{1} \bar{\Xmodeindex}_{2} \Xmodeindex'_{1} \Xmodeindex'_{2}}_{\BXTM \bar{\BXRM} \BXRM'} .
		\end{split}
	\end{equation}
\end{widetext}
The theory so far is completely general and applies to any semiconducting crystalline system.
In the next section we consider the specific case when the conduction and valence bands are spin-degenerate, which yields some simplifications.

\subsection{Spin-decoupled biexciton eigenvalue problem}

In the expressions from the previous sections, the indices $\Xmodeindex_{1}, \Xmodeindex_{2}, \dots$ are used to label all exciton states unambiguously.
It often happens that the conduction and valence bands have some degree of degeneracy due to additional quantum numbers.
This degeneracy gives rise to multiple degenerate exciton states, which must somehow be distinguished from each other within these labels.
The most common example of this is the electron and hole spin in normal semiconductors: one can label excitons via the coupled electron--hole spin or via the spin projection of each individual particle.
For hydrogenic excitons, for instance, each exciton level is four-fold degenerate, as we have one singlet state and three triplet states, or equivalently, four up/down combinations for the electron and hole.
When both electrons and holes are effectively spin-$\frac{1}{2}$ particles and the single-particle Hamiltonian is the same for both spin projections, the total exciton wave function is a product state of a spin and orbital part.
As shown in our Ref.\ \cite{noordman2025variational}, the biexciton eigenvalue problem may then be brought into a simpler form.
This is accomplished by coupling the electron spins into the spin state $\vec{S}_{\mathrm{e}} = (S_{\mathrm{e}}, S^{z}_{\mathrm{e}})$ and the hole spins into the spin state $\vec{S}_{\mathrm{h}} = (S_{\mathrm{h}}, S^{z}_{\mathrm{h}})$, where $S_{\mathrm{e}}$ ($S_{\mathrm{h}}$) and $S_{\mathrm{e}}^{z}$ ($S_{\mathrm{h}}^{z}$) are the total spin and spin projection of the coupled electrons (holes).
In this basis, the biexciton wave function transforms as
\begin{equation}
\label{eq:BXWFTransfSPinBasis}
	\Psi^{\vec{S}_{\mathrm{e}} \vec{S}_{\mathrm{h}}, \XmodeindexNoSpin_{1} \XmodeindexNoSpin_{2}}_{\BXmodeindex \BXTM}(\BXRM) = (-1)^{S_{\mathrm{e}} + S_{\mathrm{h}}} \hspace{0.3mm} \Psi^{\vec{S}_{\mathrm{e}} \vec{S}_{\mathrm{h}}, \XmodeindexNoSpin_{2} \XmodeindexNoSpin_{1}}_{\BXmodeindex \BXTM}({-}\BXRM - G \BXTM) ,
\end{equation}
where here and below the $\XmodeindexNoSpin$ indices reference the remaining \emph{spin-independent} part of the excitonic labels (for hydrogen-like excitons, this would be the principal quantum number $n$ and the orbital angular momentum quantum numbers).

This pairwise-coupled spin basis is convenient because it allows the biexciton problem to be solved independently for the various values of $S_{\mathrm{e}}$.
Indeed, in this basis we have
\begin{equation}
	\label{eq:RMatrixElSpinBasis}
	\begin{split}
		& R^{\vec{S}_{\mathrm{e}} \vec{S}_{\mathrm{h}} \vec{S}_{\mathrm{e}}' \vec{S}_{\mathrm{h}}' , \XmodeindexNoSpin_{1} \XmodeindexNoSpin_{2} \XmodeindexNoSpin_{1}' \XmodeindexNoSpin_{2}'}_{\BXTM \BXRM \BXRM'} \\
		&= \delta_{\vec{S}_{\mathrm{e}} \vec{S}_{\mathrm{e}}'} \delta_{\vec{S}_{\mathrm{h}} \vec{S}_{\mathrm{h}}'} \Big(\mathcal{V} \delta_{\BXRM \BXRM'} \delta_{\XmodeindexNoSpin_{1} \XmodeindexNoSpin_{1}'} \delta_{\XmodeindexNoSpin_{2} \XmodeindexNoSpin_{2}'} + (-1)^{S_{\mathrm{e}}} [\mathcal{K}^{\mathrm{e}}]^{\XmodeindexNoSpin_{1} \XmodeindexNoSpin_{2} \XmodeindexNoSpin_{1}' \XmodeindexNoSpin_{2}'}_{\BXTM \BXRM \BXRM'} \Big) ,
	\end{split}
\end{equation}
which is diagonal in the ingoing and outgoing coupled spins.
The part inside the parentheses depends on $S_{\mathrm{e}}$ only and will be denoted by $[R_{S_{\mathrm{e}}}]^{\smash[b]{\XmodeindexNoSpin_{1} \XmodeindexNoSpin_{2} \XmodeindexNoSpin_{1}' \XmodeindexNoSpin_{2}'}}_{\BXTM \BXRM \BXRM'}$.
The matrix element $[\mathcal{K}^{\mathrm{e}}]^{\smash[b]{\XmodeindexNoSpin_{1} \XmodeindexNoSpin_{2} \XmodeindexNoSpin_{1}' \XmodeindexNoSpin_{2}'}}_{\BXTM \BXRM \BXRM'}$ in Eq.\ \eqref{eq:RMatrixElSpinBasis} is calculated from the expression in Eq.\ \eqref{eq:Kc} with the understanding that the factorizable spin-dependent parts of the wave functions (and the corresponding spin sums) are omitted, and hence it depends on the $\XmodeindexNoSpin$ labels only.
The inverse of $R_{S_{\mathrm{e}}}$ when the sum over states is restricted again admits the Dyson series
\begin{equation}
	\begin{split}
		& [R_{S_{\mathrm{e}}}^{-1}]^{\XmodeindexNoSpin_{1} \XmodeindexNoSpin_{2} \XmodeindexNoSpin_{1}' \XmodeindexNoSpin_{2}'}_{\BXTM \BXRM \BXRM'} \\
		& = \mathcal{V} \delta_{\vec{q} \vec{q}'} \delta_{\XmodeindexNoSpin_{1} \XmodeindexNoSpin_{1}'} \delta_{\XmodeindexNoSpin_{2} \XmodeindexNoSpin_{2}'} - \frac{(-1)^{S_{\mathrm{e}}}}{\mathcal{V}} \sum_{\XmodeindexNoSpin_{3} \XmodeindexNoSpin_{4} \bar{\vec{q}}} [R_{S_{\mathrm{e}}}^{-1}]^{\XmodeindexNoSpin_{1} \XmodeindexNoSpin_{2} \XmodeindexNoSpin_{3} \XmodeindexNoSpin_{4}}_{\BXTM \BXRM \bar{\BXRM}} [\mathcal{K}^{\mathrm{e}}]^{\XmodeindexNoSpin_{3} \XmodeindexNoSpin_{4} \XmodeindexNoSpin_{1}' \XmodeindexNoSpin_{2}'}_{\BXTM \bar{\BXRM} \BXRM'} .
	\end{split}
\end{equation}
The different exciton--exciton interaction terms again depend only on $S_{\mathrm{e}}$ and $S_{\mathrm{h}}$ via
\begin{subequations}
	\begin{align}
		[\mathcal{U}^{0}]^{\vec{S}_{\mathrm{e}} \vec{S}_{\mathrm{h}} \vec{S}_{\mathrm{e}}' \vec{S}_{\mathrm{h}}' , \XmodeindexNoSpin_{1} \XmodeindexNoSpin_{2} \XmodeindexNoSpin_{1}' \XmodeindexNoSpin_{2}'}_{\BXTM \BXRM \BXRM'} &= \delta_{\vec{S}_{\mathrm{e}} \vec{S}_{\mathrm{e}}'} \delta_{\vec{S}_{\mathrm{h}} \vec{S}_{\mathrm{h}}'} [\mathcal{U}^{0}]^{\XmodeindexNoSpin_{1} \XmodeindexNoSpin_{2} \XmodeindexNoSpin_{1}' \XmodeindexNoSpin_{2}'}_{\BXTM \BXRM \BXRM'} , \\
		[\mathcal{U}^{\mathrm{e}}]^{\vec{S}_{\mathrm{e}} \vec{S}_{\mathrm{h}} \vec{S}_{\mathrm{e}}' \vec{S}_{\mathrm{h}}' , \XmodeindexNoSpin_{1} \XmodeindexNoSpin_{2} \XmodeindexNoSpin_{1}' \XmodeindexNoSpin_{2}'}_{\BXTM \BXRM \BXRM'} &= \delta_{\vec{S}_{\mathrm{e}} \vec{S}_{\mathrm{e}}'} \delta_{\vec{S}_{\mathrm{h}} \vec{S}_{\mathrm{h}}'} (-1)^{S_{\mathrm{e}} + 1} [\mathcal{U}^{\mathrm{e}}]^{\XmodeindexNoSpin_{1} \XmodeindexNoSpin_{2} \XmodeindexNoSpin_{1}' \XmodeindexNoSpin_{2}'}_{\BXTM \BXRM \BXRM'} , \\
		[\mathcal{U}^{\mathrm{h}}]^{\vec{S}_{\mathrm{e}} \vec{S}_{\mathrm{h}} \vec{S}_{\mathrm{e}}' \vec{S}_{\mathrm{h}}' , \XmodeindexNoSpin_{1} \XmodeindexNoSpin_{2} \XmodeindexNoSpin_{1}' \XmodeindexNoSpin_{2}'}_{\BXTM \BXRM \BXRM'} &= \delta_{\vec{S}_{\mathrm{e}} \vec{S}_{\mathrm{e}}'} \delta_{\vec{S}_{\mathrm{h}} \vec{S}_{\mathrm{h}}'} (-1)^{S_{\mathrm{h}} + 1} [\mathcal{U}^{\mathrm{h}}]^{\XmodeindexNoSpin_{1} \XmodeindexNoSpin_{2} \XmodeindexNoSpin_{1}' \XmodeindexNoSpin_{2}'}_{\BXTM \BXRM \BXRM'} , \\
		[\mathcal{U}^{\mathrm{X}}]^{\vec{S}_{\mathrm{e}} \vec{S}_{\mathrm{h}} \vec{S}_{\mathrm{e}}' \vec{S}_{\mathrm{h}}' , \XmodeindexNoSpin_{1} \XmodeindexNoSpin_{2} \XmodeindexNoSpin_{1}' \XmodeindexNoSpin_{2}'}_{\BXTM \BXRM \BXRM'} &= \delta_{\vec{S}_{\mathrm{e}} \vec{S}_{\mathrm{e}}'} \delta_{\vec{S}_{\mathrm{h}} \vec{S}_{\mathrm{h}}'} (-1)^{S_{\mathrm{e}} + S_{\mathrm{h}}} [\mathcal{U}^{\mathrm{X}}]^{\XmodeindexNoSpin_{1} \XmodeindexNoSpin_{2} \XmodeindexNoSpin_{1}' \XmodeindexNoSpin_{2}'}_{\BXTM \BXRM \BXRM'} ,
	\end{align}
\end{subequations}
where all matrix elements on the right-hand side are understood to be calculated from their corresponding expressions in Appendix\ \ref{app:InteractionTermsExpressions} by omitting all spin-related parts.

The expressions above are again diagonal in the pairwise-coupled spins.
As a result, the biexciton problem becomes
\begin{widetext}
	\begin{equation}
		\label{eq:2XSchrodingerEquationSc}
			\big(\varepsilon^{\XmodeindexNoSpin_{1}}_{\Gamma \BXTM + \BXRM} + \varepsilon^{\XmodeindexNoSpin_{2}}_{\bar{\Gamma} \BXTM - \BXRM}\big) \Psi^{\vec{S}_{\mathrm{e}} \vec{S}_{\mathrm{h}}, \XmodeindexNoSpin_{1} \XmodeindexNoSpin_{2}}_{\BXmodeindex \BXTM}(\BXRM) + \frac{1}{\mathcal{V}} \sum_{\XmodeindexNoSpin_{1}' \XmodeindexNoSpin_{2}' \BXRM'} [V_{S_{\mathrm{e}}}^{\text{eff}}]^{\XmodeindexNoSpin_{1} \XmodeindexNoSpin_{2} \XmodeindexNoSpin_{1}' \XmodeindexNoSpin_{2}'}_{\BXTM \BXRM \BXRM'} \Psi^{\vec{S}_{\mathrm{e}} \vec{S}_{\mathrm{h}}, \XmodeindexNoSpin_{1}' \XmodeindexNoSpin_{2}'}_{\BXmodeindex \BXTM}(\BXRM') = \mathcal{E}^{\BXmodeindex}_{\vec{Q}} \Psi^{\vec{S}_{\mathrm{e}} \vec{S}_{\mathrm{h}}, \XmodeindexNoSpin_{1} \XmodeindexNoSpin_{2}}_{\BXmodeindex \BXTM}(\BXRM) .
	\end{equation}
	with the matrix elements of the effective potential given by
	\begin{equation}
	\label{eq:XXVeffSc}
		\begin{split}
			[V^{\text{eff}}_{S_{\mathrm{e}}}]^{\XmodeindexNoSpin_{1} \XmodeindexNoSpin_{2} \XmodeindexNoSpin'_{1} \XmodeindexNoSpin'_{2}}_{\BXTM \BXRM \BXRM'} = \frac{(-1)^{S_{\mathrm{e}}}}{2 \mathcal{V}}
			&\sum_{\XmodeindexNoSpin_{3} \XmodeindexNoSpin_{4} \bar{\vec{q}}} [R_{S_{\mathrm{e}}}^{-1}]^{\XmodeindexNoSpin_{1} \XmodeindexNoSpin_{2} \XmodeindexNoSpin_{3} \XmodeindexNoSpin_{4}}_{\BXTM \BXRM \bar{\BXRM}} \big(\varepsilon^{\XmodeindexNoSpin_{3}}_{\Gamma \BXTM + \bar{\BXRM}} + \varepsilon^{\XmodeindexNoSpin_{4}}_{\bar{\Gamma} \BXTM - \bar{\BXRM}} - \varepsilon^{\XmodeindexNoSpin_{1}'}_{\Gamma \BXTM + \BXRM'} - \varepsilon^{\XmodeindexNoSpin_{2}'}_{\bar{\Gamma} \BXTM - \BXRM'}\big) [\mathcal{K}^{\mathrm{e}}]^{\XmodeindexNoSpin_{3} \XmodeindexNoSpin_{4} \XmodeindexNoSpin_{1}' \XmodeindexNoSpin_{2}'}_{\BXTM \bar{\BXRM} \BXRM'} \\
			+ \, \frac{1}{\mathcal{V}} &\sum_{\XmodeindexNoSpin_{3} \XmodeindexNoSpin_{4} \bar{\vec{q}}} [R_{S_{\mathrm{e}}}^{-1}]^{\XmodeindexNoSpin_{1} \XmodeindexNoSpin_{2} \XmodeindexNoSpin_{3} \XmodeindexNoSpin_{4}}_{\BXTM \BXRM \bar{\BXRM}} [\mathcal{U}^{0} - (-1)^{S_{\mathrm{e}}} \mathcal{U}^{\mathrm{e}}]^{\XmodeindexNoSpin_{3} \XmodeindexNoSpin_{4} \XmodeindexNoSpin_{1}' \XmodeindexNoSpin_{2}'}_{\BXTM \bar{\BXRM} \BXRM'} .
		\end{split}
	\end{equation}
\end{widetext}
In this way the problem has been reduced because not all spin states need to be considered simultaneously, but we can solve it for the different values of $S_{\mathrm{e}}$ one at a time.
The total valence spin $S_{\mathrm{h}}$ becomes a spectator variable whose correct value for a given $S_{\mathrm{e}}$ must be chosen for each individual state $\BXmodeindex$ in order to satisfy the reflection property of Eq.\ \eqref{eq:BXWFTransfSPinBasis}.
That is, solutions of Eq.\ \eqref{eq:2XSchrodingerEquationSc} with even (odd) parity under $\BXRM \rightarrow {-}\BXRM - G \BXTM$ and $\XmodeindexNoSpin_{1} \leftrightarrow \XmodeindexNoSpin_{2}$ correspond to the parity of $S_{\mathrm{h}}$ being equal (opposite) of that of $S_{\mathrm{e}}$.
Furthermore, the biexciton energies in such systems are independent of the spin projections $S_{\mathrm{e}}^{z}$ and $S_{\mathrm{h}}^{z}$.

In summary, the contents of Sec.\ \ref{sec:genBXProblem} apply to any system, while in the specific case where electrons and holes of different spins are described by the same Hamiltonian one may apply the theory of this section.
In any case the biexciton spectrum will be the same, but considering different values of $S_{\mathrm{e}}$ separately when possible reduces the computational complexity of the problem.
It is also worth mentioning that some systems may lack the spin degeneracy assumed in this section, but possess some other degenerate degree of freedom such as a valley or orbital index.
If such a pseudospin can be identified, one may perform a decoupling analogous to that of this section in the appropriate variables and reduce the full biexciton problem to several simpler ones.

In the remainder of the article we numerically apply the theory to a few simple systems to exemplify how this theory can be applied to study the effects of nontrivial band geometry on the interactions between excitons.

\section{Square-lattice tight-binding model}
\label{sec:TBEq}

Before studying models with band geometry and topology, we consider a bilayer electron--hole system with a simple tight-binding square-lattice dispersion on each layer.
The top layer contains a conduction band with dispersion $\epsilon^{c}_{\XRM}$ and the bottom layer contains a valence band with dispersion $\epsilon^{v}_{\XRM}$.
These read
\begin{subequations}
\label{eq:cosineDispersions}
	\begin{align}
		\epsilon^{c}_{\vec{k}} &= \hphantom{{-}} t (2 - \cos k_{x} - \cos k_{y}) + E_{\mathrm{g}} / 2, \\
		\epsilon^{v}_{\vec{k}} &= {-}t (2 - \cos k_{x} - \cos k_{y}) - E_{\mathrm{g}} / 2 ,
	\end{align}
\end{subequations}
where $E_{\mathrm{g}}$ is the band gap between the two layers and the hopping strength $t$ has been taken equal in each layer for simplicity.
We consider a ``contact'' interaction between the two layers, i.e., the electrons only interact with a strength $U$ when they sit at sites right on top of each other within the two layers.
The on-site repulsions within each layer are taken equal to each other and denoted by $U_{\text{rep}}$.

The BSE for this simple model reads
\begin{equation}
\label{eq:BSETightBindingSquareLatt}
	(\epsilon^{c}_{\XTM + \XRM} - \epsilon^{v}_{\XRM}) \Phi_{\XTM}(\XRM) - \frac{U}{\mathcal{V}} \sum_{\XRM'} \Phi_{\XTM}(\XRM') = \varepsilon_{\XTM} \Phi_{\XTM}(\XRM) ,
\end{equation}
where $\mathcal{V}$ is now the surface area of the 2D system and we have chosen $\gamma = 1$ without loss of generality.
We have also omitted the exciton index $\Xmodeindex$ because we will consider the ground-state excitons only.
In fact, Eq.\ \eqref{eq:BSETightBindingSquareLatt} delivers a single bound state below the continuum, whose dispersion $\varepsilon_{\XTM}$ is the single solution to the equation
\begin{equation}
\label{eq:SLTBDispEq}
	1 = \frac{U}{\mathcal{V}} \sum_{\XRM} \frac{1}{\epsilon^{c}_{\XTM + \XRM} - \epsilon^{v}_{\XRM} - \varepsilon_{\XTM}} .
\end{equation}
The wave function corresponding to this bound state reads
\begin{equation}
\label{eq:SLTBPhi}
	\Phi_{\XTM} = \frac{\mathcal{N}}{\epsilon^{c}_{\XTM + \XRM} - \epsilon^{v}_{\XRM} - \varepsilon_{\XTM}} ,
\end{equation}
where $\mathcal{N}$ is a normalization constant that depends on $t$ and $U$.

We have numerically solved Eq.\ \eqref{eq:SLTBDispEq}, as well as the BSE of Eq.\ \eqref{eq:BSETightBindingSquareLatt} on an $N \times N$ momentum grid.
We have checked that the exciton energies obtained through both methods are consistent with each other, though deviations start to occur for very lightly bound excitons with $\varepsilon_{\XTM} / \Delta_{\XTM} \gtrsim 0.999$, where $\Delta_{\XTM}$ stands for the minimal energy of an electron--hole pair with total momentum $\XTM$.
However, in this regime our approximation of neglecting the scattering states in the biexciton calculation is not justified anyway due to the small splitting between the bound state and the continuum, so we focus on the regime when they are relatively tightly bound.
The results presented here correspond to $N = 50$, and we have checked that the results are the same for $N = 20$, $30$, and $40$, indicating excellent convergence for this problem.
After obtaining the exciton wave functions we evaluate the exchange integral $\mathcal{K}^{\mathrm{e}}$ and the exciton--exciton matrix elements $\mathcal{U}^{0}$ and $\mathcal{U}^{\mathrm{e}}$ by restricting all exciton states to the ground state.
We stress that this is only a convenient approximation, but that our theory allows for the inclusion of an arbitrary number of excited states.
We set the total biexciton momentum $\BXTM$ to zero, as we expect this to correspond to the largest biexciton binding energies owing to the direct band gap between both bands.
For each value of the coupled electron spin $S_{\mathrm{e}} = 0, 1$, we evaluate the matrix $R_{S_{\mathrm{e}}}$ and the $S_{\mathrm{e}}$-dependent effective potential of Eq.\ \eqref{eq:XXVeffSc}.
We then solve the biexciton eigenvalue problem of Eq.\ \eqref{eq:2XSchrodingerEquationSc}, and find that all eigenvalues are real as expected.
In Appendix \ref{app:numerics} we briefly elaborate on the numerical details.

Only for the case $S_{\mathrm{e}} = 0$ do we find bound states of two excitons.
This can be understood by comparing them with the problem of two interacting excitons with parabolic electron and hole bands and a Coulombic $1/r$ electron--hole potential, which was considered in our previous Ref.\ \cite{noordman2025variational}.
There we explicitly demonstrated that in the limit of the hole being much more massive than the electron the effective exciton--exciton potentials corresponding to $S_{\mathrm{e}} = 0$ and $S_{\mathrm{e}} = 1$ become the Heitler--London singlet and triplet potentials, respectively.
The former contains an attractive well and supports bound states of two hydrogenic excitons.
Meanwhile, the latter is fully repulsive and forbids the formation of electron-triplet biexcitons at this level of approximation, i.e., without including corrections from excited states that provide a van der Waals-type induced dipole--dipole interaction.
In this sense, the exciton problem considered in this section is simply the tight-binding analog of two interacting hydrogenic ground-state excitons.
For this reason, we note that including the effect of excited states may introduce a shallow potential well in the $S_{\mathrm{e}} = 1$ channel and result in lightly bound biexcitons there as well.

\begin{figure}[!t]
	\centering
	\includegraphics[width=\linewidth]{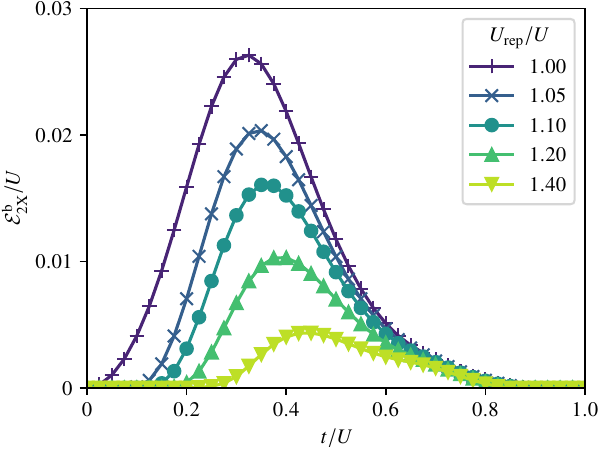}
	\caption{Biexciton binding energy $\mathcal{E}^{\mathrm{b}}_{2X}$ for $S_{\mathrm{e}} = 0$ in a simple 2D square-lattice tight-binding model with the cosine dispersions of Eq.\ \eqref{eq:cosineDispersions} as a function of the ratio $t / U$ between the hopping and interaction strengths.
	In this model, there are no bound two-exciton states with $S_{\mathrm{e}} = 1$ when considering only ground-state excitons in the biexciton problem.
	Different curves correspond to different values of the electron--electron and hole--hole repulsion strengths relative to the electron--hole attraction strength.
	For $t / U \gtrsim 0.9$ no bound biexcitons appear, and the behavior of $\mathcal{E}^{\mathrm{b}}_{2X}$ is nonmonotonic because the electron--hole attraction competes with the repulsion between identical particles.
	The calculations have been performed on a $50 \times 50$ momentum grid.}
	\label{fig:EbbiX_squareLattCos}
\end{figure}

The biexciton binding energies for the case $S_{\mathrm{e}} = 0$ are shown in Fig.\ \ref{fig:EbbiX_squareLattCos} as a function of the ratio $t / U$.
The biexciton binding energies $\mathcal{E}^{\mathrm{b}}_{2 \mathrm{X}}$ are defined as the minimum energy of two free excitons with the same total momentum (in this case 0) minus the total energy $\mathcal{E}$ of the biexciton.
The different curves correspond to different values of the ratio $U_{\text{rep}} / U$ between the repulsion strength between like particles to the electron--hole attraction strength.
We numerically find that there is a maximum value of $t / U \approx 0.9$ above which two excitons cannot bind together.
In other words, for any given $t$ there is a minimum value of the interaction strength $U$ required to produce a two-exciton bound state.
Starting from $t / U \gg 1$ and progressively increasing the interaction strength $U$ initially causes the biexciton binding to increase, as seen by moving from right to left in the figure.
Eventually, the binding energy attains a maximum value and decreases as $t / U$ becomes smaller, vanishing in the limit $t / U \rightarrow 0$.
This signals that the internal attraction between electrons and holes competes with the repulsion between like particles, leading to a nonmonotonic behavior of the biexciton binding energy with $t / U$.
Note that in this calculation the band gap $E_{\mathrm{g}}$ does not play a role, as long it is large enough to prevent the formation of an excitonic insulator.

The limit $t / U = 0$ is interesting to study analytically in this model.
For finite $U$, this implies that there are two completely flat bands.
In this case, the exciton energy is equal to $E_{\mathrm{g}} - U$ and independent of the momentum, and the exciton wave function is constant on the Brillouin zone.
It is then easy to show that the elements of the exciton--exciton interaction $\mathcal{U}$ all vanish, that is, $\mathcal{U}^{0} = \mathcal{U}^{\mathrm{e}} = \mathcal{U}^{\mathrm{h}} = \mathcal{U}^{\mathrm{X}} = 0$.
Consequently, bound biexcitons are absent in this flat-band model as the effective exciton--exciton potential of Eq.\ \eqref{eq:XXVeffSc} vanishes.
This is far from a general consequence of flat-band systems.
In the following sections, we consider excitons on flat bands with nontrivial quantum geometry.
We will see that these properties not only strongly modify the exciton spectrum with respect to the trivial case of this section, but can also have significant consequences on the biexciton spectrum.

\section{Excitons in flat-band bilayers}
\label{sec:fbbls}

A simple yet very interesting situation to test our theory of exciton--exciton interactions is that of flat bands in 2D.
We consider systems with a single flat valence band and a single flat conduction band, and assume a twofold spin degeneracy for each.
We again assume that the electrons and holes live in spatially separated bilayers with negligible wave function overlap, as this is an experimentally desirable situation due to strong tunability and lengthened exciton lifetimes.
We again consider contact interactions with strengths $U$ and $U_{\text{rep}}$ for the electron--hole attraction and the electron--electron and hole--hole repulsions, respectively, which act when the particles lie on the same in-plane coordinate in their respective layers.

Let us first prove some useful relations that hold in the case of flat bands and a contact interaction.
We begin by rearranging the exciton BSE as
\begin{equation}
    \frac{1}{\mathcal{V}} \sum_{\mathclap{c' v' \XRM'}} \, [\tilde{D} - \tilde{X}]^{cvc'v'}_{\XTM \XRM \XRM'} \Phi^{c'v'}_{\Xmodeindex \XTM \XRM'} = \bigg(\frac{E_{\mathrm{g}} - \varepsilon^{\Xmodeindex}_{\XTM}}{U}\bigg) \Phi^{cv}_{\Xmodeindex \XTM \XRM} ,
\end{equation}
where $\tilde{D}$ and $\tilde{X}$ are the direct and exchange matrix elements with $U = 1$ and $E_{\mathrm{g}}$ is the energy difference between the two flat bands.
The interaction matrix on the left-hand side is independent of $U$, and thus so are the wave functions $\Phi$.
It is then clear that the quantity $(E_{\mathrm{g}} - \varepsilon^{\Xmodeindex}_{\XTM}) / U$ remains constant with $U$.
Identifying $(E_{\mathrm{g}} - \varepsilon^{\Xmodeindex}_{\XTM})$ as the binding energy $\varepsilon^{\mathrm{b}}_{\XTM}$ of an exciton with total momentum $\XTM$ in the flat-band system, we conclude that $\varepsilon^{\mathrm{b}}_{\XTM} / U$ is independent of $U$.
Equivalently, the exciton dispersions differ only by a shift and a rescaling, namely
\begin{equation}
\label{eq:XDispUUprimeRel}
    \varepsilon_{\XTM}(U') = \frac{U'}{U} \varepsilon_{\XTM}(U) + \bigg(1 - \frac{U'}{U}\bigg) E_{\mathrm{g}} .
\end{equation}
Note that these results hold under the assumption of on-site electron--hole interactions and a single flat band of each kind, i.e., they are independent of the electron--hole overlap and thus also hold for monolayers.

Under the approximation of a single exciton mode, one can then show that a similar relation holds for the binding energy of biexcitons with zero total momentum, namely that $\mathcal{E}^{\mathrm{b}}_{2 \mathrm{X}} / U$ only depends on the ratio $U_{\text{rep}} / U$ between the intralayer repulsion and the interlayer attraction strengths.
This can be derived directly from the biexciton eigenvalue problem of Eq.\ \eqref{eq:2XSchrodingerEquation} at $\BXTM = \mathbf{0}$ as follows.
First, one restricts all exciton states $\Xmodeindex$ to the same one and subtracts the two-exciton continuum energy $\min_{\XTM} 2 \varepsilon_{\XTM}$ at both sides.
Then, using the definition of the exciton binding energy $\varepsilon^{\mathrm{b}}_{\XTM} = E_{\mathrm{g}} - \varepsilon_{\XTM}$ and using that $\varepsilon^{\mathrm{b}}_{\XTM} / U$ is independent of $U$ one straightforwardly obtains the desired result.
Note that $\mathcal{E}^{\mathrm{b}}_{2 \mathrm{X}} / U$ being a function of $U_{\text{rep}} / U$ only is contingent on the single-particle bands being flat and the use of a contact interaction in all channels.
Indeed, the fact that the exciton wave functions are independent of the electron--hole attraction strength causes the different terms of the effective exciton potential to depend on $U$ or $U_{\text{rep}}$ solely via an overall multiplicative factor.

\subsection{Imbalanced Lieb-lattice model}
\label{sec:imbalancedLieb}

\begin{figure}[!t]
    \centering
    \includegraphics[width=\linewidth]{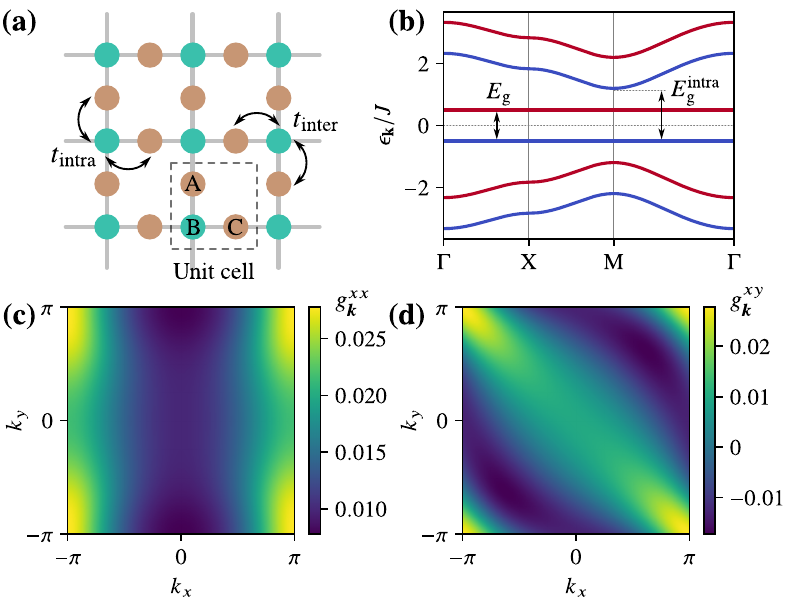}
    \caption{
    Imbalanced Lieb-lattice bilayer model and its associated band structure and quantum metric.
    The sketch in \textbf{(a)} shows the lattice that models each of the two layers in the system, together with the imbalanced intra- and intercell hopping strengths $t_{\text{intra}} = J (1 + \delta)$ and $t_{\text{inter}} = J (1 - \delta)$.
    In \textbf{(b)} we have plotted the dispersion of the three bands within the electron layer (red) and the hole layer (blue) and show the band gaps defined in the main text.
    The chemical potential lies at zero energy, so that the two flat bands are expected to host most of the excitonic weight as long as $U \ll E_{\mathrm{g}}^{\text{intra}}$.
    Finally, \textbf{(c)} and \textbf{(d)} show the quantum-metric components $g^{xx}_{\vec{k}}$ and $g^{xy}_{\vec{k}}$, respectively.
    We have set $\delta = 0.6$ everywhere.
    }
    \label{fig:model_LiebJdelta}
\end{figure}

We begin by considering the square Lieb-lattice bilayer system of Ref.\ \cite{ying2024flat}.
Each layer has three orbitals per site denoted by A, B, and C.
The intracell hopping strength $t_{\text{intra}} = J (1 + \delta)$ is taken to be different from the intercell hopping strength $t_{\text{inter}} = J (1 - \delta)$, where $J$ sets the hopping energy scale and $\delta \in [0, 1]$ quantifies this mismatch.
Due to the latter, we call this the imbalanced Lieb-lattice model, and we note that it could in principle be realized with a designer lattice where the distance between two A and B atoms within the same unit cell is smaller than the distance between two A and B atoms belonging to different unit cells (and the same for two B and C atoms).
The Bloch Hamiltonian in the aforementioned orbital basis reads
\begin{equation}
    H_{0, \XRM} =
    \begin{bmatrix}
        0 & a_{\XRM} & 0 \\
        a_{\XRM}^{*} & 0 & b^{}_{\XRM} \\
        0 & b_{\XRM}^{*} & 0
    \end{bmatrix} ,
\end{equation}
with
\begin{subequations}
    \begin{align}
        a_{\XRM} &= {-} J (1 + \delta) - J (1 - \delta) \ec^{\iu k_{y}} , \\
        b_{\XRM} &= {-} J (1 + \delta) - J (1 - \delta) \ec^{\iu k_{x}} .
    \end{align}
\end{subequations}
For simplicity we set $\delta$ and $J$ equal in both layers and also set the lattice constant to $1$.
A sketch of the model and its band structure are shown in Fig.\ \ref{fig:model_LiebJdelta}.
For $\delta > 0$ there is an isolated flat band, which is separated from the other dispersive bands in the same layer by the intralayer energy gap $E_{\mathrm{g}}^{\text{intra}} = 2 \sqrt{2} \delta J$ shown in the figure.
Also, in accordance with our treatment above we assume that the interaction is the same for all orbitals, $U_{\mathrm{AA}} = U_{\mathrm{BB}} = U_{\mathrm{CC}} \equiv U$, and only acts when two orbitals are placed exactly on top of each other, meaning that $U_{\mathrm{AB}} = U_{\mathrm{BC}} = U_{\mathrm{AC}} = 0$.

We are interested in the case where the excitons are formed with electrons and holes residing on the flat bands only, i.e., there is negligible excitonic weight on the dispersive bands.
For this reason we assume that the layers are gated in such a way that the chemical potential lies exactly between the two flat bands.
We then disregard the effect of the dispersive bands by assuming that the electron-hole interaction strength $U$ is much smaller than the intralayer energy gap $E^{\text{intra}}_{\mathrm{g}}$.
In this way the exciton problem becomes completely independent of the hopping scale $J$, which is a particularity of the fact that we consider perfectly flat bands.
The interaction strength $U$ is then the only relevant energy scale in the problem and we accordingly choose it as our unit of energy.

\begin{figure}[!t]
    \centering
    \includegraphics[width=\linewidth]{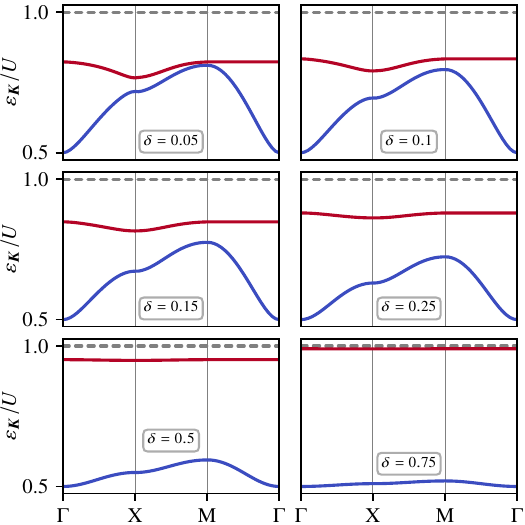}
    \caption{Exciton dispersions in the square Lieb-lattice model for different values of $\delta$.
    The ground-state exciton, shown in blue, has a finite effective mass at the $\Gamma$ point despite the flatness of the electron and hole bands.
    The two exciton modes are well separated in energy for $\delta \gtrsim 0.1$.
    The grey dashed line marks the electron--hole continuum, which is constant for the flat-band system.
    The axes are equal in all plots.
    }
    \label{fig:FIG_Xspec_LiebJdelta}
\end{figure}

\begin{figure}[!t]
    \centering
    \includegraphics[width=\linewidth]{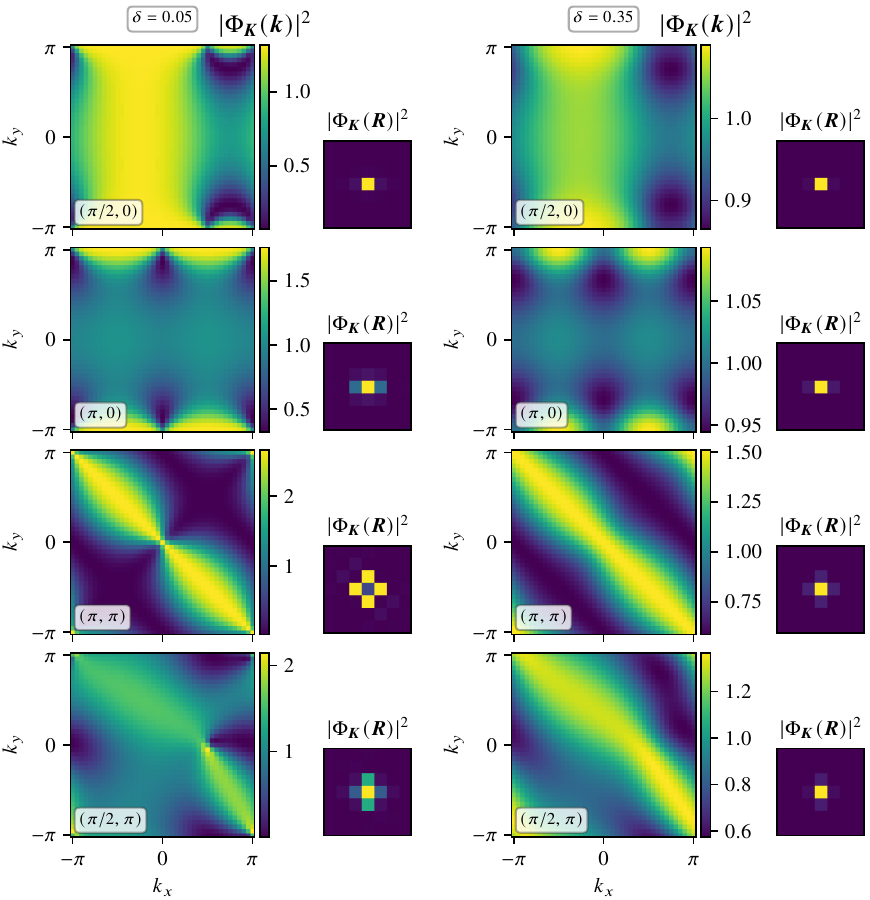}
    \caption{Momentum-space probability densities of excitonic states in the square Lieb-lattice model for two different values of $\delta$ indicated at the top.
    The four plots in each column correspond to different total exciton momenta $\XTM$, indicated on the bottom left of each plot.
    The small plots on the side correspond to the position-space probability densities, obtained via Fourier transformation of the momentum-space wave functions.
    Note that the momenta are implicitly expressed in units of the lattice constant and that $\Phi_{\XTM}(\vec{R})$ is defined on the Bravais lattice, so that each square in these plots represents the excitonic weight on one unit cell.
    }
    \label{fig:FIG_WFs_LiebJdelta}
\end{figure}

We numerically solve the single-exciton BSE within subspace spanned by the flat bands on a $40 \times 40$ momentum grid, where the convergence has been checked against smaller grid sizes.
Note that the exchange interaction vanishes in our setup because the electron and hole wave functions do not overlap in real space.
The exciton dispersions are shown in Fig.\ \ref{fig:FIG_Xspec_LiebJdelta} for different values of $\delta$.
There are two exciton modes below the continuum threshold, which are well separated in energy everywhere in the Brillouin zone unless $\delta \ll 1$.
Despite the fact that the single-particle bands are completely flat, the ground-state excitons have a finite mass around the $\Gamma$ point and are dispersive.
In Fig.\ \ref{fig:FIG_WFs_LiebJdelta} we show the probability density of the excitonic ground state for two different values of $\delta$.

This finite exciton mass can be traced back to the existence of a nontrivial quantum metric on the flat band.
The quantum geometric tensor of the latter is given by
\begin{equation}
\label{eq:QGTLieb}
    \mathcal{Q}^{\mu \nu}_{\XRM} = \langle \partial_{\mu} u_{\XRM} \vert (1 - \vert u_{\XRM} \rangle \langle u_{\XRM} \vert) \vert \partial_{\nu} u_{\XRM} \rangle ,
\end{equation}
where $\partial_{\mu}$ stands for the partial derivative with respect to $k_{\mu}$ and
\begin{equation}
    u^{}_{\XRM} = \frac{[{-} b_{\XRM}^{}, 0, a_{\XRM}^{*}]^{\mathsf{T}}}{\sqrt{|a_{\vec{k}}|^{2} + |b_{\vec{k}}|^{2}}}
\end{equation}
is the eigenfunction corresponding to the flat band.
The quantum metric is defined as the real part of Eq.\ \eqref{eq:QGTLieb} and its components are given by
\begin{subequations}
    \begin{align}
        g^{xx}_{\vec{k}} &= \frac{\delta_{-}^{2} (1 + \delta^{2} + \delta_{-} \delta_{+} \cos k_{y})}{2 [2 (1 + \delta^{2}) + \delta_{-} \delta_{+} (\cos k_{x} + \cos k_{y})]^{2}} , \\
        g^{yy}_{\vec{k}} &= \frac{\delta_{-}^{2} (1 + \delta^{2} + \delta_{-} \delta_{+} \cos k_{x})}{2 [2 (1 + \delta^{2}) + \delta_{-} \delta_{+} (\cos k_{x} + \cos k_{y})]^{2}} , \\
        g^{xy}_{\vec{k}} &= \frac{\delta_{-}^{2} [\delta_{-}^{2} + \delta_{-} \delta_{+} (\cos k_{x} + \cos k_{y}) + \delta_{+}^{2} \cos(k_{x} + k_{y})]}{4 [2 (1 + \delta^{2}) + \delta_{-} \delta_{+} (\cos k_{x} + \cos k_{y})]^{2}} ,
    \end{align}
\end{subequations}
with $\delta_{\pm} \equiv 1 \pm \delta$.
These are plotted in Fig.\ \ref{fig:model_LiebJdelta}.
As shown in Ref.\ \cite{ying2024flat}, the inverse-mass tensor for the ground-state exciton has components \footnote{While this should be accompanied by a factor of $(2 \pi)^{2} / \mathcal{A}_{\text{BZ}}$, with $\mathcal{A}_{\text{BZ}}$ the area of the Brillouin zone, in our case this is 1 because we have set the lattice constant to 1.}
\begin{equation}
    [\excitonMass^{-1}]^{}_{\mu \nu} = U \int \frac{\mathrm{d}^{2} \XRMmagnitude}{(2 \pi)^{2}} \, g^{\mu \nu}_{\XRM} ,
\end{equation}
where the integral is performed over the Brillouin zone.
This shows that a nontrivial quantum metric gives a finite exciton mass.
Meanwhile, if $g$ vanishes then $\excitonMass \rightarrow \infty$ and the exciton bands will be flat, which is precisely what we found in the square-lattice tight-binding model of Sec.\ \ref{sec:TBEq}.
In this model this corresponds to $\delta = 1$.

\begin{figure}[!t]
    \centering
    \includegraphics[width=\linewidth]{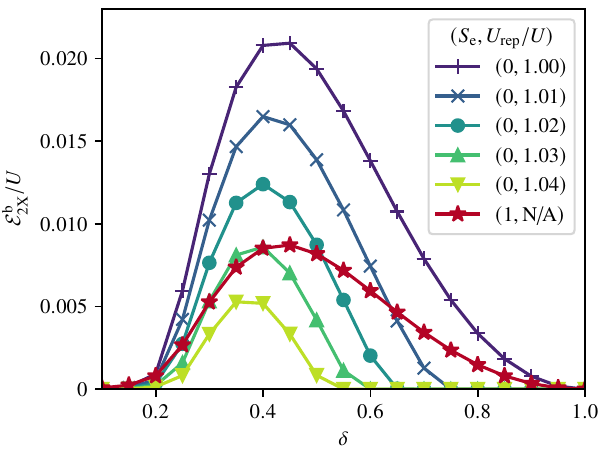}
    \caption{Biexciton binding energies $\mathcal{E}^{\mathrm{b}}_{2 \mathrm{X}}$ in the imbalanced Lieb-lattice model as a function of the dimensionless parameter $\delta$ that quantifies the quantum metric of the flat band.
    The nontrivial quantum metric not only strongly influences the binding energies, but is also able to produce biexciton states in the electronic spin-triplet configuration, which contrasts with the Heitler--London result in conventional systems at this level of approximation.
    In addition, the binding energy of biexcitons with $S_{\mathrm{e}} = 1$ is independent of the electron--electron repulsion strength $U_{\text{rep}}$, making them very stable compared to their counterparts with $S_{\mathrm{e}} = 0$.
    All calculations have been performed on a $40 \times 40$ momentum grid.
    }
    \label{fig:EbbiX_LiebJdelta}
\end{figure}

Next we consider the biexciton problem, for which we keep only the excitonic ground state.
As seen from the dispersions, this approximation is good everywhere except in the case of very small $\delta$.
We assume that the model is spin-degenerate, so that both spin projections for the electron and the hole are described by their own copy of the Lieb-lattice model described above.
With this in mind we evaluate the spin-dependent exciton--exciton potential of Eq.\ \eqref{eq:XXVeffSc} restricted to the excitonic ground state and solve the biexciton Schr\"odinger equation.
Figure\ \ref{fig:EbbiX_LiebJdelta} shows the biexciton binding energies as a function of the parameter $\delta$ that parametrizes the quantum metric across the Brillouin zone.

It is clear that bound biexcitons exist in a large interval of $\delta$, and we also see that their binding energies depend very strongly and nonmonotonically on this parameter.
The existence of bound two-exciton states is a direct consequence of the nontrivial quantum metric, as in the square-lattice model of the previous section we found no such states in the limit of zero hopping.
Furthermore, the biexciton binding energies for the singlet configuration $S_{\mathrm{e}} = 0$ decrease very rapidly with increasing intralayer repulsion.
This is similar to the square-lattice tight-binding result of Fig.\ \ref{fig:EbbiX_squareLattCos}, except that the decrease is much steeper.
On the other hand, we observe that there now exist bound biexcitons in the triplet configuration $S_{\mathrm{e}} = 1$.
This is in stark contrast with the Heitler--London limit of hydrogenic excitons and with the square-lattice model of Sec.\ \ref{sec:TBEq} and is again a direct consequence of the nontrivial quantum metric.

Numerically, we have found that the triplet biexcitons do not depend on the repulsion strength $U_{\text{rep}}$.
This can be understood analytically by defining a ``repulsive'' part of the exciton--exciton interaction via $\mathcal{U}^{\text{rep}}_{S_{\mathrm{e}}} \equiv (\mathcal{U}^{0}_{\mathrm{ee}} + \mathcal{U}^{0}_{\mathrm{hh}}) - (-1)^{S_{\mathrm{e}}} (\mathcal{U}^{\mathrm{e}}_{\mathrm{ee}} + \mathcal{U}^{\mathrm{e}}_{\mathrm{hh}})$, where the name comes from the fact that it is the part of the interaction that is proportional to $U_{\text{rep}}$.
We then find that $\mathcal{U}^{\text{rep}}_{S_{\mathrm{e}}}$ is odd for $S_{\mathrm{e}} = 1$, that is, $[\mathcal{U}^{\text{rep}}_{S_{\mathrm{e}} = 1}]^{}_{\BXRM, {-}\BXRM'} = {-}[\mathcal{U}^{\text{rep}}_{S_{\mathrm{e}} = 1}]^{}_{\BXRM \BXRM'}$.
Because the ground-state biexciton wave function is even, we have $\sum_{\BXRM'} [\mathcal{U}^{\text{rep}}_{S_{\mathrm{e}} = 1}]^{}_{\BXRM \BXRM'} \Psi^{}_{\BXRM'} = 0$.
This means that the ground-state biexciton binding energy for $S_{\mathrm{e}} = 1$ is independent of the repulsion strength $U_{\text{rep}}$.
More fundamentally, that this has to be the case follows from the fact that the physics can never depend on an on-site repulsion strength between particles of equal spin due to the Pauli exclusion principle.
As a result, triplet biexcitons are very stable against strong intralayer repulsion of identical particles and can exist even when their $S_{\mathrm{e}} = 0$ counterparts cannot.
The situation is thus completely reversed with respect to the Heitler--London case.

\subsection{Square-lattice $\boldsymbol{\zeta}$ model}

We also consider a model with nontrivial quantum metric given by \cite{hofmann2023superconductivity,verma2024geometric}
\begin{equation}
\label{eq:H0kzeta}
    H_{0, \vec{k}} = t (2 - p_{\vec{k}}) \sigma_{0} + t' \cos (\zeta p_{\vec{k}}) \sigma_{x} + t' \sin (\zeta p_{\vec{k}}) \sigma_{y} ,
\end{equation}
where $p_{\vec{k}} \equiv \cos k_{x} + \cos k_{y}$.
In this model $\zeta$ quantifies the quantum geometric tensor via 
\begin{equation}
    \mathcal{Q}^{\mu \nu}_{\vec{k}} = \frac{\zeta^{2}}{4} \sin k_{\mu} \sin k_{\nu} ,
\end{equation}
which is fully real because of inversion and time-reversal symmetries.
Therefore, there is no Berry curvature and the bands have a zero Chern number, but the quantum metric is nontrivial and equal to $\mathcal{Q}^{\mu \nu}_{\vec{k}}$.
The eigenenergies are given by $\epsilon_{\vec{k}}^{\pm} = t (2 - p_{\vec{k}}) \pm |t'|$ and have the convenient property that they are independent of $\zeta$.
As such, similarly to the Lieb-lattice model, we can tune the quantum geometry without affecting the dispersion of the bands of interest.
We will set $t = 0$, which makes the bands fully flat and allows us to make use of the properties discussed above.

\begin{figure}[!t]
    \centering
    \includegraphics[width=\linewidth]{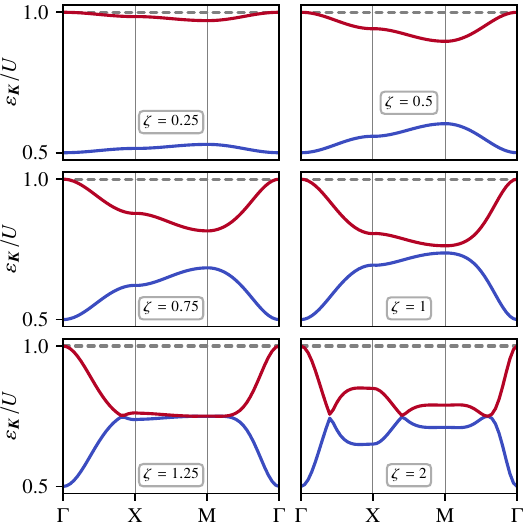}
    \caption{Exciton dispersions in the square-lattice $\zeta$ model for different values of $\zeta$.
    Again, the ground-state exciton has a finite effective mass at the $\Gamma$ and for $\zeta \lesssim 1.2$ the two exciton modes are separated in energy.
    For larger $\zeta$, the dispersions cross and give rise to multiple Dirac cones within the Brillouin zone.
    The grey dashed line marks the electron--hole continuum and the axes are equal in all plots.
    }
    \label{fig:FIG_Xspec_squareLattZeta}
\end{figure}

\begin{figure}[!t]
    \centering
    \includegraphics[width=\linewidth]{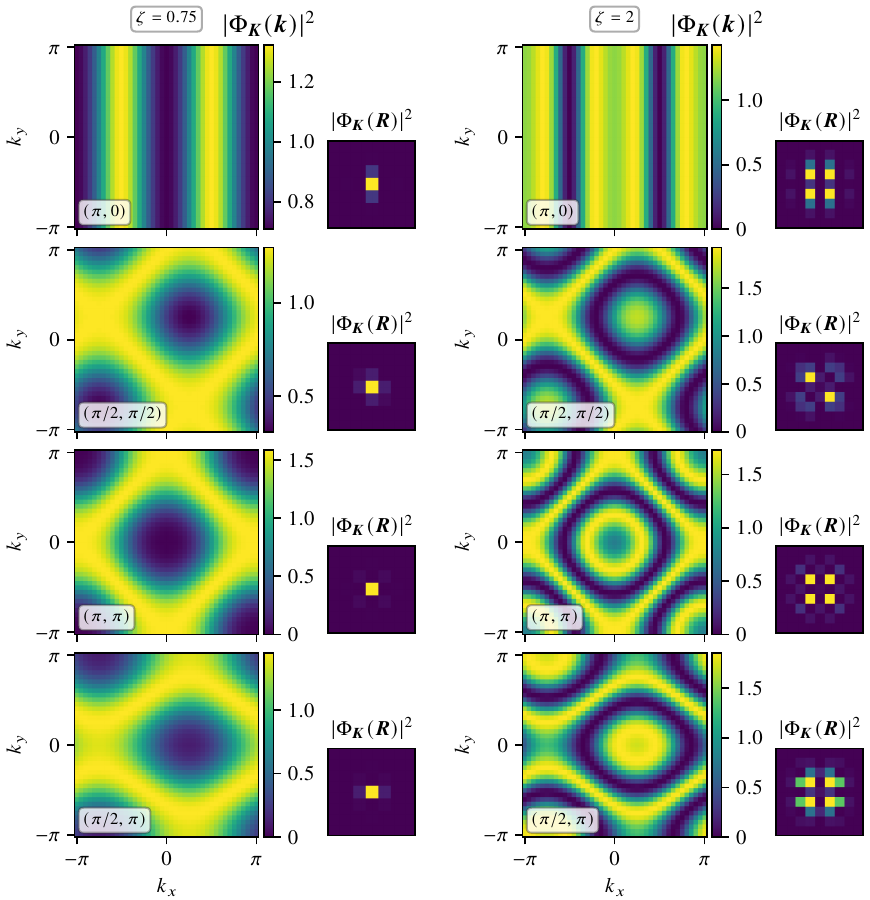}
    \caption{Momentum- and position-space probability densities of excitonic states in the square-lattice $\zeta$ model for two different values of $\zeta$ indicated at the top.
    The same comments under Fig.\ \ref{fig:FIG_WFs_LiebJdelta} apply.
    }
    \label{fig:FIG_WFs_squareLattZeta}
\end{figure}

Even though in this model both bands are flat, we still consider a gated bilayer system in order to avoid including the exchange interaction, and include only the two bands closest to the Fermi level.
Again, the interaction is assumed to be diagonal in the two ``orbitals'' of the model in Eq.\ \eqref{eq:H0kzeta} with orbital-independent strengths $U$ and $U_{\text{rep}}$.
We set $\zeta$ equal in both layers for simplicity and show the resulting exciton spectrum in Fig.\ \ref{fig:FIG_Xspec_squareLattZeta} for several values of $\zeta$.
There are again two bound electron--hole states which for $\zeta \leq 1.2$ are separated in energy.
The ground-state exciton always has the $\Gamma$-point energy $\varepsilon_{\Gamma} / U = 0.5$ and again shows a finite effective mass around this point.
Interestingly, for larger values of $\zeta$ the dispersions of the two states cross and give rise to multiple excitonic Dirac points within the Brillouin zone at the same energy, namely $\varepsilon_{\vec{K}} / U = 0.75$.

\begin{figure}[!t]
    \centering
    \includegraphics[width=\linewidth]{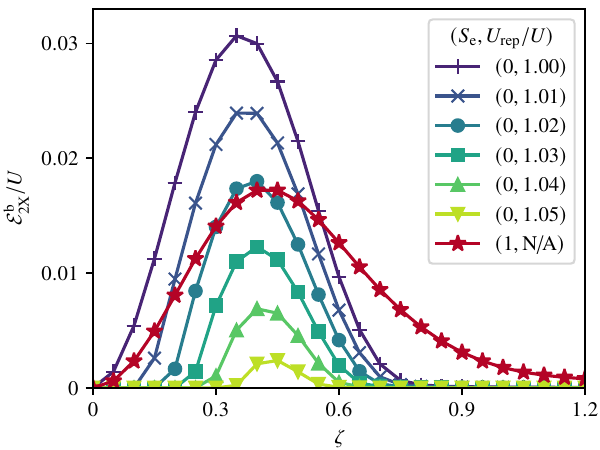}
    \caption{Biexciton binding energies $\mathcal{E}^{\mathrm{b}}_{2X}$ in the square-lattice $\zeta$ model as a function of the dimensionless parameter $\zeta$ that quantifies the quantum metric of the flat band.
    We find the same physics as in the imbalanced Lieb-lattice model of Fig.\ \ref{fig:EbbiX_LiebJdelta}.
    The calculations have again been performed on a $40 \times 40$ momentum grid.
    }
    \label{fig:FIG_EbbiX_squareLattZeta}
\end{figure}

We subsequently compute the effective exciton--exciton potential for the ground-state excitons and show some associated probability densities in Fig.\ \ref{fig:FIG_WFs_squareLattZeta}.
We solve the biexciton problem in the spin-degenerate case and plot the binding energies in Fig.\ \ref{fig:FIG_EbbiX_squareLattZeta}.
We do this in the single-exciton approximation whereby we only consider the ground-state excitons, which is valid in the regime $\zeta \lesssim 1.2$ in which the two exciton modes are gapped.
Even though the model is microscopically very different from that in Sec.\ \ref{sec:imbalancedLieb}, our conclusions still hold.
Namely, that the nontrivial quantum metric enables bound triplet biexcitons which are very stable against strong electron--electron repulsion within the same layer.
The results in this section highlight the significant effect of quantum geometry on exciton--exciton interactions and related phenomena.

\section{Summary and outlook}
\label{sec:chXXint:summary}

In this work we have presented a theory of exciton--exciton interactions in general crystalline semiconductors that takes into account the full internal structure of the excitons.
Specifically, we have derived a variational eigenvalue problem for the bound state of two excitons from where an effective exciton--exciton potential matrix could be identified.
We have then evaluated this potential and solved the associated biexciton eigenvalue problem in some simple tight-binding models with a contact interaction.
An important conclusion is that a nontrivial quantum metric can stabilize triplet biexcitons even at the lowest level of approximation, which contrasts with the simple hydrogenic case where the triplet potential is fully repulsive.

The effective biexciton eigenvalue equation can be applied to the study of biexciton spectra in theoretically and experimentally relevant material platforms.
Indeed, our theory naturally incorporates the hybridization and quantum-geometric effects of the underlying single-particle bands and is thus applicable to systems with arbitrary single-particle Hamiltonians.
Particularly interesting examples are topological insulators, transition-metal dichalcogenides, and moiré heterostructures.
This paves the way for understanding biexciton formation and related effects in the presence of topology, quantum geometry, and twisting effects.
Our preliminary results already show that in the simplest case of tight-binding models with a contact interaction, quantum geometry can strongly influence the biexcitonic spectrum.
The generality of our theory makes it possible to systematically study the effects of band topology, geometry, and Berry curvature on the biexciton spectrum of systems with specific symmetries and in various spatial dimensions.
In the same way that a general classification of topological excitons may be achieved in the coming years, it would also be interesting to consider the topological properties of biexcitons and how these are related to those of the individual constituents.

Beyond the free biexciton picture, our theory can be used as a basis for quantitative calculations of correlated excitonic phases.
In Ref.\ \cite{maisel2026correlated} we studied the Bose--Einstein condensation of excitons on moat bands, but used a toy interaction between excitons based on a simplistic point-charge model.
With knowledge of the underlying single-particle model, this could be improved by instead using the exciton--exciton interaction $\mathcal{U}$ of Eq.\ \eqref{eq:XXintUAllTerms} as an approximation to the bare potential to be renormalized via the $T$-matrix procedure.
On another note, Ref.\ \cite{gotting2022moire} derives an excitonic moiré Bose--Hubbard model for interlayer excitons in twisted transition-metal dichalcogenide bilayers.
Their calculation of the repulsion strength between two sites also employs a toy interaction between excitons, which again could be refined by instead using the matrix elements $\mathcal{U}$ via
\begin{equation}
    U_{\vec{R} \vec{R}'} = \int \mathrm{d}^{2} r \, \mathrm{d}^{2} r' \, |w_{\vec{R}}(\vec{r})|^{2} \hspace{0.45mm} \mathcal{U}_{\vec{R} - \vec{R}'}(\vec{r}, \vec{r}') |w_{\vec{R}'}(\vec{r}')|^{2} .
\end{equation}
Here, $w_{\vec{R}}(\vec{r})$ are excitonic Wannier functions, and $\mathcal{U}_{\vec{R} - \vec{R}'}(\vec{r}, \vec{r}')$ is the coordinate-space counterpart of $\mathcal{U}_{\BXTM \BXRM \BXRM'}$, where in particular the $\BXTM$ label is transformed via wannierization analogously to the excitonic wave functions \cite{haber2023maximally}.
In this case, the form factors that dress the electrostatic interaction in the direct and exchange matrix elements arise from the diagonalization of the moiré Hamiltonian.
The ``flavors'' $\genSpinOne, \genSpinTwo, \dots$ we have used throughout will then correspond to the moiré reciprocal vectors $\vec{G}$, as the moiré Hamiltonian in a plane-wave representation has the schematic structure \cite{bistritzer2011moire,wu2018hubbard}
\begin{equation}
    [H^{\text{moiré}}_{0,\vec{k}}]^{}_{\vec{G} \vec{G}'} = \frac{\hbar^{2} (\vec{k} + \vec{G})^{2}}{2 m^{*}} \delta_{\vec{G} \vec{G}'} + V^{\text{moiré}}_{\vec{G} - \vec{G}'} .
\end{equation}
Here we have omitted possible spin and layer indices to highlight the status of the reciprocal vectors as basis elements instead of a conserved label.
Just like in the quantum-geometric models, these form factors will introduce into the BSE all the information about the underlying moiré potential and we expect a significant effect on the exciton--exciton interaction matrix elements, especially in the regime of very flat bands.
Note that in the moiré case we need the generalization to the case of nonconserved electronic flavors at the electrostatic vertex provided in Appendix\ \ref{app:generalVertex}.

Overall, our theory is highly general and versatile and we hope it will be applied to further theoretical and numerical studies and to explain experimental observations of exciton--exciton interactions and related phenomena.

\section{Acknowledgments}

This work is supported by the Delta-ITP consortium, which is part of the Netherlands Organisation for Scientific Research (NWO).
We also acknowledge the research program ``Materials for the Quantum Age'' (QuMat) for financial support.
This program (registration number 024.005.006) is part of the Gravitation program financed by the Dutch Ministry of Education, Culture and Science (OCW).

\appendix

\section{Explicit expressions}
\label{app:InteractionTermsExpressions}

\noindent
Here we provide the explicit expressions for the different terms that appear in the expression for the total exciton--exciton interaction of Eq.\ \eqref{eq:XXintUAllTerms}.
Two excitons in the states $(\Xmodeindex_{1}', \Xmodeindex_{2}')$ with momenta $(\Gamma \BXTM + \BXRM', \bar{\Gamma} \BXTM - \BXRM')$ scatter into another exciton pair in the states $(\Xmodeindex_{1}, \Xmodeindex_{2})$ with momenta $(\Gamma \BXTM + \BXRM, \bar{\Gamma} \BXTM - \BXRM)$.
Of course, the total biexciton momentum $\vec{Q}$ is conserved.

We omit the exciton and momentum labels on the left-hand side of all equations for the sake of brevity, i.e., we write $\mathcal{U}^{t}_{d}$ in place of $[\mathcal{U}^{t}_{d}]^{\smash[b]{\Xmodeindex_{1} \Xmodeindex_{2} \Xmodeindex_{\smash[t]{1}}' \Xmodeindex_{\smash[t]{2}}'}}_{\BXTM \BXRM \BXRM'}$ for each exchange type $t$ and diagram $d$.
Furthermore, all terms shown below are individually Hermitian, i.e., $[\mathcal{U}^{t}_{d}]^{\smash[b]{\Xmodeindex_{1} \Xmodeindex_{2} \Xmodeindex_{\smash[t]{1}}' \Xmodeindex_{\smash[t]{2}}'}}_{\BXTM \BXRM \BXRM'} = \big([\mathcal{U}^{t}_{d}]^{\smash[b]{\Xmodeindex_{1}' \Xmodeindex_{2}' \Xmodeindex_{1} \Xmodeindex_{2}}}_{\BXTM \BXRM' \BXRM}\big)^{*}$ for all $t$ and $d$ as they apply.
In each case we show both the diagram and its algebraic expression, with time flowing from left to right.
Arrows pointing to the right represent electrons, while arrows pointing to the left represent holes (which are equivalent to valence electrons flowing backwards).
We recall that $\bar{\gamma} = 1 - \gamma$, $\bar{\Gamma} = 1 - \Gamma$, $g = 2 \gamma - 1$, $G = 2 \Gamma - 1$, with $\gamma$ and $\Gamma$ the arbitrary numbers introduced in the definition of the exciton and biexciton creation operators of Eqs.\ \eqref{eq:XCreationOpGamma} and \eqref{eq:biexcitonCreationOperatorGamma}, respectively.
Finally, we note that if the underlying electrostatic vertex does not conserve the flavor labels $i$ and $j$, the form factors within these expressions should be replaced by their generalized version according to the explanation in Appendix\ \ref{app:generalVertex}.

\begin{widetext}

\subsection{Direct interaction $\boldsymbol{\mathcal{U}^{0}}$}

	\noindent
	These diagrams include all processes where one of the particles of an exciton scatters against one of the particles of the other exciton, with no exchange of particles taking place.
	The total $\mathcal{U}^{0}$ is the sum of the following terms:

	\begin{itemize}
		\item Direct electron--electron scattering: 
		\begin{figure*}[!h]
			\centering
			\includegraphics[width=0.42\textwidth]{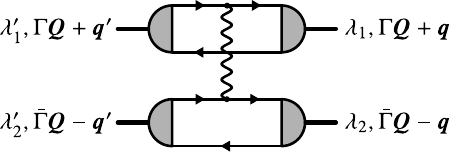}
		\end{figure*}
		\begin{equation}
		\label{eq:U0ccScattering}
			\begin{split}
				\mathcal{U}^{0}_{\text{ee}} = \frac{1}{\mathcal{V}^{2}} \sum_{\XRM \XRM'} \sum_{\genSpinOne \genSpinTwo} \sum_{\text{bands}}
				[\Phi^{cv}_{\Xmodeindex_{1}, \Gamma \BXTM + \BXRM}(\XRM)]^{*}
				\Phi^{c'v}_{\Xmodeindex_{1}',\Gamma \BXTM + \BXRM'}(\XRM - \bar{\gamma}(\BXRM - \BXRM'))
				[\Phi^{\bar{c}\bar{v}}_{\Xmodeindex_{2}, \bar{\Gamma} \BXTM - \BXRM}(\XRM')]^{*}
				\Phi^{\bar{c}'\bar{v}}_{\Xmodeindex_{2}', \bar{\Gamma} \BXTM - \BXRM'}(\XRM' + \bar{\gamma}(\BXRM - \BXRM'))& \\
				\times \, V^{\genSpinOne \genSpinTwo}_{\BXRM - \BXRM'}
				\big\langle u^{c}_{\XRM + \gamma (\Gamma \BXTM + \BXRM)} \big\vert \genSpinOne \big\rangle
				\big \langle \genSpinOne \big\vert u^{c'}_{\XRM - \bar{\gamma} \BXRM + \BXRM' + \gamma \Gamma \BXTM} \big\rangle
				\big\langle u^{\bar{c}}_{\XRM' + \gamma (\bar{\Gamma} \BXTM - \BXRM)} \big\vert \genSpinTwo \big\rangle
				\big \langle \genSpinTwo \big\vert u^{\bar{c}'}_{\XRM' + \bar{\gamma} \BXRM - \BXRM' + \gamma \bar{\Gamma} \BXTM} \big\rangle& ,
			\end{split}
		\end{equation}
		\item Direct hole--hole scattering: 
		\begin{figure*}[!h]
			\centering
			\includegraphics[width=0.42\textwidth]{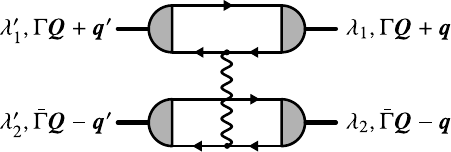}
		\end{figure*}
		\begin{equation}
		\label{eq:U0vvScattering}
			\begin{split}
				\mathcal{U}^{0}_{\text{hh}} = \frac{1}{\mathcal{V}^{2}} \sum_{\XRM \XRM'} \sum_{\genSpinOne \genSpinTwo} \sum_{\text{bands}}
				[\Phi^{cv}_{\Xmodeindex_{1}, \Gamma \BXTM + \BXRM}(\XRM)]^{*}
				\Phi^{cv'}_{\Xmodeindex_{1}', \Gamma \BXTM + \BXRM'}(\XRM + \gamma (\BXRM - \BXRM'))
				[\Phi^{\bar{c}\bar{v}}_{\Xmodeindex_{2}, \bar{\Gamma} \BXTM - \BXRM}(\XRM')]^{*}
				\Phi^{\bar{c}\bar{v}'}_{\Xmodeindex_{2}', \bar{\Gamma} \BXTM - \BXRM'}(\XRM' - \gamma (\BXRM - \BXRM'))& \\
				\times \, V^{\genSpinOne \genSpinTwo}_{\BXRM - \BXRM'}
				\big\langle u^{v'}_{\XRM + \gamma \BXRM - \BXRM' - \bar{\gamma} \Gamma \BXTM} \big\vert \genSpinOne \big\rangle
				\big \langle \genSpinOne \big\vert u^{v}_{\XRM - \bar{\gamma} (\Gamma \BXTM + \BXRM)} \big\rangle
				\big\langle u^{\bar{v}'}_{\XRM' - \gamma \BXRM + \BXRM' - \bar{\gamma} \bar{\Gamma} \BXTM} \big\vert \genSpinTwo \big\rangle
				\big \langle \genSpinTwo \big\vert u^{\bar{v}}_{\XRM' - \bar{\gamma} (\bar{\Gamma} \BXTM - \BXRM)} \big\rangle& , 
			\end{split}
		\end{equation}
		\item Direct electron--hole scattering: 
		\begin{figure*}[!h]
			\centering
			\includegraphics[width=0.42\textwidth]{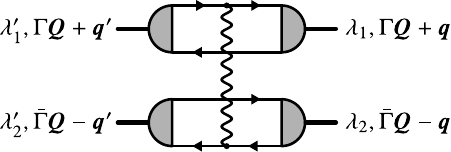}
		\end{figure*}
		\begin{equation}
		\label{eq:U0cvScattering}
			\begin{split}
				\mathcal{U}^{0}_{\text{eh}} = {-} \frac{1}{\mathcal{V}^{2}} \sum_{\XRM \XRM'} \sum_{\genSpinOne \genSpinTwo} \sum_{\text{bands}}
				[\Phi^{cv}_{\Xmodeindex_{1}, \Gamma \BXTM + \BXRM}(\XRM)]^{*}
				\Phi^{c'v}_{\Xmodeindex_{1}', \Gamma \BXTM + \BXRM'}(\XRM - \bar{\gamma} (\BXRM - \BXRM'))
				[\Phi^{\bar{c}\bar{v}}_{\Xmodeindex_{2}, \bar{\Gamma} \BXTM - \BXRM}(\XRM')]^{*}
				\Phi^{\bar{c}v'}_{\Xmodeindex_{2}', \bar{\Gamma} \BXTM - \BXRM'}(\XRM' - \gamma (\BXRM - \BXRM'))& \\
				\times \, \Big(
				V^{\genSpinOne \genSpinTwo}_{\BXRM - \BXRM'}
				\big\langle u^{c}_{\XRM + \gamma (\Gamma \BXTM + \BXRM)} \big\vert \genSpinOne \big\rangle
				\big \langle \genSpinOne \big\vert u^{c'}_{\XRM - \bar{\gamma} \BXRM + \BXRM' + \gamma \Gamma \BXTM} \big\rangle
				\big\langle u^{v'}_{\XRM' - \gamma \BXRM + \BXRM' - \bar{\gamma} \bar{\Gamma} \BXTM}  \big\vert \genSpinTwo \big\rangle
				\big \langle \genSpinTwo \big\vert u^{\bar{v}}_{\XRM' - \bar{\gamma} (\bar{\Gamma} \BXTM - \BXRM)} \big\rangle& \\
				- \, V^{\genSpinOne \genSpinTwo}_{\XRM - \XRM' + g \BXRM + (g \Gamma + \bar{\gamma}) \BXTM}
				\big\langle u^{c}_{\XRM + \gamma (\Gamma \BXTM + \BXRM)} \big\vert \genSpinOne \big\rangle
				\big \langle \genSpinOne \big\vert u^{\bar{v}}_{\XRM' - \bar{\gamma} (\bar{\Gamma} \BXTM - \BXRM)} \big\rangle
				\big\langle u^{v'}_{\XRM' - \gamma \BXRM + \BXRM' - \bar{\gamma} \bar{\Gamma} \BXTM}  \big\vert \genSpinTwo \big\rangle
				\big \langle \genSpinTwo \big\vert u^{c'}_{\XRM - \bar{\gamma} \BXRM + \BXRM' + \gamma \Gamma \BXTM} \big\rangle& \Big) ,
			\end{split}
		\end{equation}
		\item Direct hole--electron scattering: 
		\begin{figure*}[!h]
			\centering
			\includegraphics[width=0.42\textwidth]{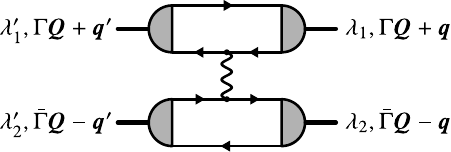}
		\end{figure*}
		\begin{equation}
		\label{eq:U0vcScattering}
			\begin{split}
				\mathcal{U}^{0}_{\text{he}} = {-} \frac{1}{\mathcal{V}^{2}} \sum_{\XRM \XRM'} \sum_{\genSpinOne \genSpinTwo} \sum_{\text{bands}}
				[\Phi^{cv}_{\Xmodeindex_{1}, \Gamma \BXTM + \BXRM}(\XRM)]^{*}
				\Phi^{cv'}_{\Xmodeindex_{1}', \Gamma \BXTM + \BXRM'}(\XRM + \gamma (\BXRM - \BXRM'))
				[\Phi^{\bar{c}\bar{v}}_{\Xmodeindex_{2}, \bar{\Gamma} \BXTM - \BXRM}(\XRM')]^{*}
				\Phi^{c'\bar{v}}_{\Xmodeindex_{2}', \bar{\Gamma} \BXTM - \BXRM'}(\XRM' + \bar{\gamma} (\BXRM - \BXRM'))& \\
				\times \, \Big(
				V^{\genSpinOne \genSpinTwo}_{\BXRM - \BXRM'}
				\big\langle u^{v'}_{\XRM + \gamma \BXRM - \BXRM' - \bar{\gamma} \Gamma \BXTM} \big\vert \genSpinOne \big\rangle
				\big \langle \genSpinOne \big\vert u^{v}_{\XRM - \bar{\gamma}(\Gamma \BXTM + \BXRM)} \big\rangle
				\big\langle u^{\bar{c}}_{\XRM' + \gamma (\bar{\Gamma} \BXTM - \BXRM)} \big\vert \genSpinTwo \big\rangle
				\big \langle \genSpinTwo \big\vert u^{c'}_{\XRM' + \bar{\gamma} \BXRM - \BXRM' + \gamma \bar{\Gamma} \BXTM} \big\rangle& \\
				- \, V^{\genSpinOne \genSpinTwo}_{\XRM - \XRM' + g \BXRM + (g \Gamma - \gamma) \BXTM}
				\big\langle u^{v'}_{\XRM + \gamma \BXRM - \BXRM' - \bar{\gamma} \Gamma \BXTM} \big\vert \genSpinOne \big\rangle
				\big \langle \genSpinOne \big\vert u^{c'}_{\XRM' + \bar{\gamma} \BXRM - \BXRM' + \gamma \bar{\Gamma} \BXTM} \big\rangle
				\big\langle u^{\bar{c}}_{\XRM' + \gamma (\bar{\Gamma} \BXTM - \BXRM)} \big\vert \genSpinTwo \big\rangle
				\big \langle \genSpinTwo \big\vert u^{v}_{\XRM - \bar{\gamma}(\Gamma \BXTM + \BXRM)} \big\rangle& \Big) . 
			\end{split}
		\end{equation}
	\end{itemize}
	
	\subsection{Electron-exchange interaction $\boldsymbol{\mathcal{U}^{\mathrm{e}}}$}

	\noindent
	These diagrams include all processes where the two excitons exchange one of their electrons.
	We recall the definition $\Delta^{cv}_{\XTM \XRM} = \epsilon^{c}_{\XRM + \gamma \XTM} - \epsilon^{v}_{\XRM - \bar{\gamma} \XTM}$.
	The total $\mathcal{U}^{\mathrm{e}}$ is the sum of the following three terms:

	\begin{itemize}
	\item Electron exchange with electron--electron scattering: 
	\begin{figure*}[!h]
		\centering
		\includegraphics[width=0.42\textwidth]{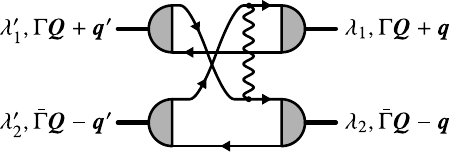}
	\end{figure*}
	\begin{equation}
		\label{eq:UCccScattering}
			\begin{split}
				\mathcal{U}^{\mathrm{e}}_{\text{ee}} = {-} \frac{1}{\mathcal{V}^{2}} \sum_{\XRM \XRM'} \sum_{\genSpinOne \genSpinTwo} \sum_{\text{bands}}
				[\Phi^{cv}_{\Xmodeindex_{1}, \Gamma \BXTM + \BXRM}(\XRM)]^{*}
				\Phi^{c'v}_{\Xmodeindex_{1}',\Gamma \BXTM + \BXRM'}(\XRM - \bar{\gamma}(\BXRM - \BXRM'))
				[\Phi^{\bar{c}\bar{v}}_{\Xmodeindex_{2}, \bar{\Gamma} \BXTM - \BXRM}(\XRM')]^{*}
				\Phi^{\bar{c}'\bar{v}}_{\Xmodeindex_{2}', \bar{\Gamma} \BXTM - \BXRM'}(\XRM' + \bar{\gamma}(\BXRM - \BXRM'))& \\
				\times \, V^{\genSpinOne \genSpinTwo}_{\XRM - \XRM' + g \BXRM + \BXRM' + \gamma G \BXTM}
				\big\langle u^{c}_{\XRM + \gamma (\Gamma \BXTM + \BXRM)} \big\vert \genSpinOne \big\rangle
				\big \langle \genSpinOne \big\vert u^{\bar{c}'}_{\XRM' + \bar{\gamma} \BXRM - \BXRM' + \gamma \bar{\Gamma} \BXTM} \big\rangle
				\big\langle u^{\bar{c}}_{\XRM' + \gamma (\bar{\Gamma} \BXTM - \BXRM)} \big\vert \genSpinTwo \big\rangle
				\big \langle \genSpinTwo \big\vert u^{c'}_{\XRM - \bar{\gamma} \BXRM + \BXRM' + \gamma \Gamma \BXTM} \big\rangle& ,
			\end{split}
		\end{equation}
	\item Electron exchange with hole--hole scattering: 
	\begin{figure*}[!h]
		\centering
		\includegraphics[width=0.42\textwidth]{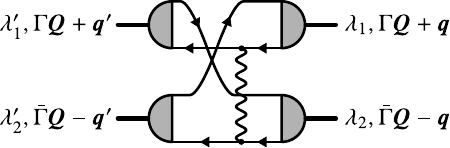}
	\end{figure*}
	\begin{equation}
	\label{eq:UCvvScattering}
		\begin{split}
			\mathcal{U}^{\mathrm{e}}_{\text{hh}} = {-} \frac{1}{\mathcal{V}^{2}} \sum_{\XRM \XRM'} \sum_{\genSpinOne \genSpinTwo} \sum_{\text{bands}}
			[\Phi^{cv}_{\Xmodeindex_{1}, \Gamma \BXTM + \BXRM}(\XRM)]^{*}
			\Phi^{cv'}_{\Xmodeindex_{2}', \bar{\Gamma} \BXTM - \BXRM'}(\XRM + \gamma (\BXRM + \BXRM' + G \BXTM))
			[\Phi^{\bar{c}\bar{v}}_{\Xmodeindex_{2}, \bar{\Gamma} \BXTM - \BXRM}(\XRM')]^{*}
			\Phi^{\bar{c}\bar{v}'}_{\Xmodeindex_{1}', \Gamma \BXTM + \XRM'}(\XRM' - \gamma (\BXRM + \BXRM' + G \BXTM))& \\
			\times \, V^{\genSpinOne \genSpinTwo}_{\XRM - \XRM' + g \BXRM + \BXRM' + \gamma G \BXTM}
			\big\langle u^{\bar{v}'}_{\XRM' - \gamma \BXRM - \BXRM' + (\gamma \bar{\Gamma} - \Gamma) \BXTM} \big\vert \genSpinOne \big\rangle
			\big \langle \genSpinOne \big\vert u^{v}_{\XRM - \bar{\gamma} (\Gamma \BXTM + \BXRM)} \big\rangle
			\big\langle u^{v'}_{\XRM + \gamma \BXRM + \BXRM' + (\gamma \Gamma - \bar{\Gamma}) \BXTM} \big\vert \genSpinTwo \big\rangle
			\big \langle \genSpinTwo \big\vert u^{\bar{v}}_{\XRM' - \bar{\gamma} (\bar{\Gamma} \BXTM - \BXRM)} \big\rangle& .
		\end{split}
	\end{equation}
	\item Electron exchange without scattering: 
	\begin{figure*}[!h]
		\centering
		\includegraphics[width=0.42\textwidth]{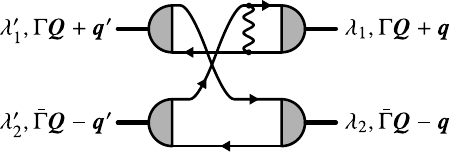}
	\end{figure*}
	\begin{equation}
	\label{eq:UCNoScattering}
		\begin{split}
			&\mathcal{U}^{\mathrm{e}}_{\times} = \frac{1}{\mathcal{V}} \sum_{\vec{k}} \sum_{\text{bands}}
			[\Phi^{cv}_{\Xmodeindex_{1}, \Gamma \BXTM + \BXRM}(\vec{k})]^{*}
			[\Phi^{c'v'}_{\Xmodeindex_{2}, \bar{\Gamma} \BXTM - \BXRM}(\XRM + g \BXRM + \BXRM' + \gamma G \BXTM)]^{*}
			\Phi^{c'v}_{\Xmodeindex_{1}', \Gamma \BXTM + \BXRM'}(\XRM - \bar{\gamma}(\BXRM - \BXRM'))
			\Phi^{cv'}_{\Xmodeindex_{2}', \bar{\Gamma} \BXTM - \BXRM'}(\XRM + \gamma (\BXRM + \BXRM' + G \BXTM)) \\
			&\times \frac{1}{2} \Big[
			\big(\Delta^{cv}_{\Gamma \BXTM + \BXRM, \XRM} - 
			\varepsilon^{\Xmodeindex_{1}}_{\Gamma \BXTM + \BXRM}\big) + 
			\big(\Delta^{c'v'}_{\bar{\Gamma} \BXTM - \BXRM, \XRM + g \BXRM + \BXRM' + \gamma G \BXTM} - 
			\varepsilon^{\Xmodeindex_{2}}_{\bar{\Gamma} \BXTM - \vec{q}}\big) + 
			\big(\Delta^{c'v}_{\Gamma \BXTM + \BXRM', \XRM - \bar{\gamma}(\BXRM - \BXRM')} - 
			\varepsilon^{\Xmodeindex_{1}'}_{\Gamma \BXTM + \BXRM'}\big) + 
			\big(\Delta^{cv'}_{\bar{\Gamma} \BXTM - \BXRM', \XRM + \gamma (\BXRM + \BXRM' + G \BXTM)} - 
			\varepsilon^{\Xmodeindex_{2}'}_{\bar{\Gamma} \BXTM - \BXRM'}\big)\Big] .
		\end{split}
	\end{equation}
	Note that the name ``without scattering'' only refers to the fact that the algebraic expression does not \emph{explicitly} contain the electrostatic interaction $V$.
	In reality, this amplitude contains all combinations of electron--hole scatterings, but they can be reduced to the above form by employing the excitonic BSE via the schematic relation $(D - X) \Phi = (E_{\mathrm{g}} - \varepsilon) \Phi$.
	To see why this happens, we observe that the diagram in principle is supposed to represent the scattering between the hole of $\Xmodeindex_{1}'$ and the electron of $\Xmodeindex_{2}'$.
	However, because an electron exchange takes place, the interaction can be ``pulled`` through the diagram so that it connects the electron and hole of $\Xmodeindex_{1}$.
	This implies that the BSE for exciton $\Xmodeindex_{1}$ can be used.
	Technically speaking, the diagram above only represents the first of the four terms in the second line of Eq.\ \eqref{eq:UCNoScattering}.
	There exist another three similar diagrams with scatterings between $(\mathrm{e}_{1'}, \mathrm{h}_{2'})$, $(\mathrm{e}_{2}, \mathrm{h}_{1})$, and $(\mathrm{e_{1}}, \mathrm{h}_{2})$.
	By similarly pulling the interactions through the diagram, these allow for the use of the BSE pertaining to the exciton modes $\Xmodeindex_{2}$, $\Xmodeindex_{1}'$, and $\Xmodeindex_{2}'$, respectively.
	This ultimately results in the four energy terms in the above expression.
	\end{itemize}

\end{widetext}
We do not also give the explicit expressions for $\mathcal{U}^{\mathrm{X}}$ and $\mathcal{U}^{\mathrm{h}}$ because their components are related to those of $\mathcal{U}^{0}$ and $\mathcal{U}^{\mathrm{e}}$ as follows:
\begin{subequations}
\label{eq:Urelations}
	\begin{align}
	\label{eq:U0UXrelation}
		[\mathcal{U}^{\mathrm{X}}_{s}]^{\Xmodeindex_{1} \Xmodeindex_{2} \Xmodeindex_{1}' \Xmodeindex_{2}'}_{\BXTM \BXRM \BXRM'}
		&= [\mathcal{U}^{0}_{s}]^{\Xmodeindex_{1} \Xmodeindex_{2} \Xmodeindex_{2}' \Xmodeindex_{1}' }_{\BXTM \BXRM, {-}\BXRM' - G \BXTM}
		= [\mathcal{U}^{\mathrm{0}}_{s}]^{\Xmodeindex_{2} \Xmodeindex_{1} \Xmodeindex_{1}' \Xmodeindex_{2}'}_{\BXTM, {-}\BXRM - G \BXTM, \BXRM'} , \\
	\label{eq:UvUcrelation}
		[\mathcal{U}^{\mathrm{h}}_{s'}]^{\Xmodeindex_{1} \Xmodeindex_{2} \Xmodeindex_{1}' \Xmodeindex_{2}'}_{\BXTM \BXRM \BXRM'}
		&= [\mathcal{U}^{\mathrm{e}}_{s'}]^{\Xmodeindex_{1} \Xmodeindex_{2} \Xmodeindex_{2}' \Xmodeindex_{1}' }_{\BXTM \BXRM, {-}\BXRM' - G \BXTM}
		= [\mathcal{U}^{\mathrm{e}}_{s'}]^{\Xmodeindex_{2} \Xmodeindex_{1} \Xmodeindex_{1}' \Xmodeindex_{2}'}_{\BXTM, {-}\BXRM - G \BXTM, \BXRM'} ,
	\end{align}
\end{subequations}
where $s \in \{\mathrm{ee}, \mathrm{hh}, \mathrm{eh}, \mathrm{he}\}$ and $s' \in \{\mathrm{ee}, \mathrm{hh}, \times\}$ label the type of scattering.
The diagrams corresponding to $\mathcal{U}^{\mathrm{X}}$ and $\mathcal{U}^{\mathrm{h}}$ are obtained by simply exchanging the labels of the ingoing excitons in $\mathcal{U}^{0}$ and $\mathcal{U}^{\mathrm{e}}$ respectively, that is, $(\Xmodeindex_{1}', \Gamma \BXTM + \BXRM') \leftrightarrow (\Xmodeindex_{2}', \bar{\Gamma} \BXTM - \BXRM')$.
Note that this exhange of momenta is equivalent to the replacement $\BXRM' \rightarrow {-} \BXRM' - G \BXTM$, as $G = 2 \Gamma - 1$.
Coupled with the biexciton wave function property $\Psi^{\Xmodeindex_{1} \Xmodeindex_{2}}_{\vec{Q}}(\BXRM) = \Psi^{\Xmodeindex_{2} \Xmodeindex_{1}}_{\vec{Q}}({-} \BXRM - G \BXTM)$, Eqs.\ \eqref{eq:U0UXrelation} and \eqref{eq:UvUcrelation} greatly simplify the biexciton eigenvalue problem.

As a consistency check we note that the combination $\mathcal{U}^{0} + \mathcal{U}^{\mathrm{e}} + \mathcal{U}^{\mathrm{h}} + \mathcal{U}^{\mathrm{X}}$, which makes up the total interaction, does not depend on the diagonal elements of the on-site repulsion interaction, i.e., on the values of $V_{\genSpinOne \genSpinOne}(\vec{R} = \vec{0})$.
That this has to be the case follows from the Pauli exclusion principle, as there cannot be two identical particles in the same state at the same site.
To see this we note that the on-site repulsion strengths $U_{\genSpinOne \genSpinOne}$ enter the momentum-space interaction $V^{\genSpinOne \genSpinTwo}_{\vec{k}}$ as an additive, $\vec{k}$-independent term.
It is then easy to see from the expressions for the different interaction components that the $U_{\genSpinOne \genSpinOne}$-dependent components of $\mathcal{U}^{0}_{\mathrm{ee}}$ cancel against those of $\mathcal{U}^{\mathrm{e}}_{\mathrm{ee}}$.
The same thing happens for the pairs $(\mathcal{U}^{0}_{\mathrm{hh}}, \mathcal{U}^{\mathrm{h}}_{\mathrm{hh}})$, $(\mathcal{U}^{\mathrm{e}}_{\mathrm{hh}}, \mathcal{U}^{\mathrm{X}}_{\mathrm{h}})$, and $(\mathcal{U}^{\mathrm{h}}_{\mathrm{ee}}, \mathcal{U}^{\mathrm{X}}_{\mathrm{ee}})$, and within each of the components $\mathcal{U}^{0}_{\mathrm{eh}}$, $\mathcal{U}^{0}_{\mathrm{he}}$, $\mathcal{U}^{\mathrm{X}}_{\mathrm{eh}}$, and $\mathcal{U}^{\mathrm{X}}_{\mathrm{he}}$ individually.
Consequently, no dependence on $U_{\genSpinOne \genSpinOne}$ exists.
Furthermore, in the biexciton eigenproblem we reduce the combination $\mathcal{U}^{0} + \mathcal{U}^{\mathrm{e}} + \mathcal{U}^{\mathrm{h}} + \mathcal{U}^{\mathrm{X}}$ to just $\mathcal{U}^{0} + \mathcal{U}^{\mathrm{e}}$ by making use of the symmetries of the wave function.
It then may seem as though the aforementioned cancellation of $U_{\genSpinOne \genSpinOne}$ is spoiled since $\mathcal{U}^{\mathrm{h}}_{\mathrm{hh}}$ and $\mathcal{U}^{\mathrm{X}}_{\mathrm{hh}}$ are not present to cancel the $U_{\genSpinOne \genSpinOne}$-dependent parts of $\mathcal{U}^{0}_{\mathrm{hh}}$ and $\mathcal{U}^{\mathrm{e}}_{\mathrm{hh}}$, respectively.
However, this is not necessary, because in the eigenproblem we may shift $\BXRM' \rightarrow {-} \BXRM' - G \BXTM$ and switch $\Xmodeindex_{1}' \leftrightarrow \Xmodeindex_{2}'$ in the summation over internal indices.
Doing this in the $\mathcal{U}^{\mathrm{e}}_{\mathrm{hh}}$ term, we see that the $U_{\genSpinOne \genSpinOne}$-dependent piece cancels against the one contained in $\mathcal{U}^{0}_{\mathrm{hh}}$, thus the whole problem remains independent of these matrix elements as expected.

\section{General electrostatic vertex}
\label{app:generalVertex}

The expressions given above are all derived in the assumption that the electrosatic vertex keeps the flavor labels $\genSpinOne$ and $\genSpinTwo$ unchanged.
However, the theory can be straightforwardly generalized to the case when these are not conserved, i.e., when the electrons scatter from $(\genSpinOne', \genSpinTwo')$ to $(\genSpinOne, \genSpinTwo)$.
In this case, the Coulomb interaction is written as $V^{\genSpinOne \genSpinTwo \genSpinOne' \genSpinTwo'}_{\vec{q}}$ and the direct and exchange matrix elements of Eq.\ \eqref{eq:DXboth} must be updated in the obvious manner.
Furthermore, the form-factor sums of the interaction matrix elements of Appendix\ \ref{app:InteractionTermsExpressions} must be generalized.
In the conserved case, they all have the structure
\begin{equation}
	\sum_{\genSpinOne \genSpinTwo} V^{\genSpinOne \genSpinTwo}_{\XRM - \XRM'} \langle u^{\alpha}_{\XRM} \vert \genSpinOne \rangle \langle \genSpinOne \vert u^{\alpha'}_{\XRM'} \rangle \langle u^{\beta}_{\BXRM} \vert \genSpinTwo \rangle \langle \genSpinTwo \vert u^{\beta'}_{\BXRM'} \rangle ,
\end{equation}
with $\XRM - \XRM' = \BXRM' - \BXRM$.
In the nonconserved case, this straightforwardly generalizes to
\begin{equation}
	\sum_{\genSpinOne \genSpinTwo \genSpinOne' \genSpinTwo'} V^{\genSpinOne \genSpinTwo \genSpinOne' \genSpinTwo'}_{\XRM - \XRM'} \langle u^{\alpha}_{\XRM} \vert \genSpinOne \rangle \langle \genSpinOne' \vert u^{\alpha'}_{\XRM'} \rangle \langle u^{\beta}_{\BXRM} \vert \genSpinTwo \rangle \langle \genSpinTwo' \vert u^{\beta'}_{\BXRM'} \rangle .
\end{equation}
The rest of the theory remains unchanged, and with this generalization it is applicable to any crystalline system no matter the nature of the electrostatic interaction.

Note that these changes are in particular important in moiré systems, as the Coulomb interaction transfers momentum and thus one must account different ingoing and outgoing moiré momenta at the same vertex.
The labels $\genSpinOne, \genSpinTwo, \dots$ now contain reciprocal momenta $\vec{G}_{\genSpinOne}, \vec{G}_{\genSpinTwo}, \dots$, respectively, and the electrostatic interaction (in the absence of any other degrees of freedom) is
\begin{equation}
	V^{\genSpinOne \genSpinTwo \genSpinOne' \genSpinTwo'}_{\vec{q}} = V_{\vec{q} + \vec{G}_{\genSpinOne} - \vec{G}_{\genSpinOne'}} \delta_{\vec{G}_{\genSpinOne} + \vec{G}_{\genSpinTwo}, \vec{G}_{\genSpinOne'} + \vec{G}_{\genSpinTwo'}}.
\end{equation}
This is especially relevant in the case of a strong moiré potential, as in this case the eigenvectors $\vert u^{\alpha}_{\XRM} \rangle$ have significant support on several moiré momenta.

\section{Numerical solution of the biexciton eigenvalue problem}
\label{app:numerics}

In this Appendix we briefly clarify the procedure followed to obtain the biexciton binding energies by solving the eigenvalue problem $h \Psi = \mathcal{E} R \Psi$.
In the case $S_{\mathrm{e}} = 1$, the matrix $R$ can be badly conditioned.
However, both $h$ and $R$ are Hermitian, and in all the cases we have considered $R$ was positive definite.
In this case the eigenvalues must be real and the problem can be solved via Cholesky decomposition, which avoids explicitly computing the inverse of $R$.
For completeness, we have also solved the problem in the form $R^{-1} h \Psi = \mathcal{E} \Psi$.
The results are identical to those obtained via Cholesky decomposition, indicating that the bad conditioning of $R$ does not pose an obstacle in the calculation of biexciton spectra.

To ensure that the results are accurate we have performed several checks, including the per-pair backward error $\Vert h \Psi - \mathcal{E} R \Psi \Vert_{2} / [(\Vert h \Vert_{\mathrm{F}} + \vert \mathcal{E} \vert \Vert R \Vert_{\mathrm{F}}) \Vert \Psi \Vert_{2}]$ for each eigenmode $\Psi$, where $\Vert \cdot \Vert_{2}$ is the 2-norm and $\Vert \cdot \Vert_{\mathrm{F}}$ is the Frobenius norm, and the $R$-orthonormality $\Vert X^{\dagger} R X - 2 \mathbb{I} \Vert_{\mathrm{F}}$, where the columns of the matrix $X$ contain all the eigenmodes $\Psi$ and the factor of $2$ comes from the fact that the individual modes are orthonormalized as $\textstyle{\frac{1}{2}} \Psi^{\dagger} R \Psi = 1$.
All residual checks produce numbers of the order of $10^{-13}$ or smaller, signaling that the results are trustworthy.

\bibliography{ref}

\end{document}